\documentclass[aps,prb,10pt,twocolumn,superscriptaddress,longbibliography,floatfix]{revtex4-2}
\usepackage{amsmath,amssymb,amsthm}
\usepackage{float}
\usepackage{placeins}   % keep floats within their major manuscript sections
\usepackage{needspace}
\usepackage{graphicx}
\usepackage{capt-of}
\graphicspath{{Manuscript_Revised_assets/figures/new_figure/}}
\usepackage[colorlinks=true,urlcolor=blue,linkcolor=blue,citecolor=blue]{hyperref}
\usepackage[dvipsnames]{xcolor}
\usepackage{bm}
\usepackage{soul}
\usepackage{pifont}
\usepackage{mathtools}
\usepackage{tikz}
\usetikzlibrary{tikzmark,arrows.meta,positioning,shapes.geometric,calc}
\usepackage{array}
\usepackage[normalem]{ulem}
\usepackage{mdframed}
\usepackage{enumitem}
\usepackage{makecell}
\usepackage{physics}
\usepackage{comment}

\newcommand{\ii}{\mathrm{i}}
\newcommand{\mc}[1]{\mathcal{#1}}
\newcommand{\mr}[1]{\mathrm{#1}}

\newcommand{\bs}[1]{\boldsymbol{#1}}

\newcommand{\Id}{\mathbf{1}}
\newcommand{\bhrho}{\overline{\hat \rho}}

\newcommand{\bfr}{\boldsymbol r}
\newcommand{\bk}{\boldsymbol k}
\newcommand{\kett}[1]{\ket{#1}\!\rangle}
\newcommand{\braa}[1]{\langle\!\bra{#1}}
\newcommand{\CI}{\kett{\mr{CI}}}
\newcommand{\fSWAP}{\operatorname{fSWAP}}
\newcommand{\rec}{\mathbf m}   % measurement record

\input{revtex_float_placement.tex}

\begin{document}
\raggedbottom

\title{Chiral Domain-Wall Modes in Topological Adaptive Quantum Dynamics}
\author{Asadullah Bhuiyan}
\affiliation{Department of Physics, Cornell University, Ithaca, New York 14853, USA}
\author{Haining Pan}
\affiliation{Department of Physics, University of Florida, Gainesville, Florida 32611, USA}
\author{Chao-Ming Jian}
\affiliation{Department of Physics, Cornell University, Ithaca, New York 14853, USA}
\date{\today}

% ------------------------------- ABSTRACT -----------------------------
\begin{abstract}
Quantum dynamical systems driven by both unitary evolution and measurements open a new landscape of nonequilibrium phases of matter. The role of topology in enriching this landscape remains less understood than that of entanglement and symmetry. A particularly interesting question concerns protected dynamical modes localized on spacetime domain walls between topologically distinct dynamical phases. These dynamical topological domain-wall modes (DTDMs) are well understood only for $0+1$D walls in $1+1$D fermionic dynamics. In this work, we study $1+1$D topological domain walls in $2+1$D charge-conserving free-fermion dynamics, modeled by topological adaptive circuits that steer the two sides of each wall toward a Chern insulator and a trivial insulator through measurements and feedforward corrections. Although the circuits target no particular wall state, each wall hosts $1+1$D chiral DTDMs. At long times in each quantum trajectory, they exhibit logarithmic scaling of entanglement and charge variance and algebraically decaying density correlations, surprisingly consistent with the ground state of a $1+1$D chiral free-fermion conformal field theory. This contrasts with purely $1+1$D charge-conserving free-fermion dynamics, which exhibits only area-law entanglement at finite measurement rates. We also show that the chiral DTDMs exhibit slow purification dynamics. When the system starts from a maximally mixed state, the DTDMs' entropy decays algebraically, accompanied by a vanishing Lyapunov gap, whereas each domain purifies on an $\mathcal{O}(1)$ timescale. Finally, we propose a physical picture for the long-time behavior of the DTDMs by treating the adaptive circuit model as a quantum cooling algorithm. Our results show that topological phases of adaptive dynamics can protect chiral modes on spacetime domain walls against repeated wavefunction collapse and Born-rule randomness.
\end{abstract}

\maketitle

\setcounter{tocdepth}{2}
\tableofcontents

% ------------------------------ SECTION II ----------------------------

\section{Introduction}
Nonequilibrium quantum dynamics combining unitary evolution and measurements is an emerging field at the interface of quantum many-body physics and quantum information. This field is motivated by rapidly advancing capabilities for controlled unitary operations and mid-circuit measurements on programmable quantum platforms. It offers new insights into the dynamical control of entangled many-body wavefunctions for quantum simulation and error correction. At the same time, studying universal quantum dynamical behavior from the complementary perspectives of quantum matter and quantum information reveals a rich landscape of nonequilibrium quantum phases of matter.

Since the seminal discovery of measurement-induced transitions between dynamical phases with different entanglement scaling~\cite{li2018quantum,SkinnerRuhmanNahum2019,fisher2023random,potter2022entanglement}, this landscape has been greatly enriched by studies of the interplay between symmetry and entanglement in dynamical systems (see, for example, Refs.~\cite{SangHsieh2021,JianBauerKeselmanLudwig2022,jian2023measurementinduced,agrawal2022entanglement,barratt2022field,lang2020entanglement,majidy2023critical,BaoChoiAltman2021SymmetryEnriched,PoboikoEtAl2023,fava2023nonlinear,LavasaniAlaviradBarkeshli2021,aziz2026measurementonly} for discrete and continuous symmetries). By comparison, how topology controls the propagation of entanglement and further enriches dynamical phases remains less systematically understood. Although nonequilibrium topological phases have been demonstrated in qubit-based dynamical systems~\cite{lavasani2021topological,LavasaniAlaviradBarkeshli2021,BaoChoiAltman2021SymmetryEnriched}, a more systematic approach has been developed for fermionic systems. Reference~\cite{JianBauerKeselmanLudwig2022} showed that free-fermion dynamics combining unitary evolution and measurements can be classified according to the Altland--Zirnbauer tenfold symmetry classes. Within each symmetry class, area-law-entangled free-fermion dynamical phases can be further divided into distinct topological classes~\cite{BhuiyanPanJian2026,PanShapourianJian2025,Oshima2025,XiaoKawabata2026}. A manifestation of the topology of the spacetime dynamics is the topology of the free-fermion states reached at long times~\cite{BhuiyanPanJian2026}. Building on this connection, Ref.~\cite{BhuiyanPanJian2026} developed a general framework of topological adaptive circuits that realize these dynamical topological phases by steering the system toward target free-fermion topological states.

With this topological classification and these circuit models in place, a natural question is what protected degrees of freedom reside on spacetime domain walls separating distinct topological dynamical phases. This question extends the study of protected modes at domain walls between equilibrium topological phases to dynamics far from equilibrium. As an example, Ref.~\cite{PanShapourianJian2025} found that a $0+1$D topological domain wall in a dynamical Majorana chain hosts a dynamical topological domain-wall mode (DTDM): a dynamical Majorana mode that carries protected entanglement despite repeated wavefunction collapses caused by measurements in its neighborhood. Similar phenomena were reported in Refs.~\cite{Oshima2025,XiaoKawabata2026}. These examples motivate the investigation of higher-dimensional DTDMs, the focus of this work.

Here, we study DTDMs in $2+1$D charge-conserving topological free-fermion dynamics. In the absence of domain walls, the topology of the relevant dynamical phase is reflected in the Chern number $\mc C$ of the area-law-entangled free-fermion states toward which the dynamics steers the system. Reference~\cite{BhuiyanPanJian2026} provided concrete topological adaptive circuit models that realize such phases. Building on these models, we construct circuit dynamics in which two topologically distinct dynamical phases, targeting a Chern insulator (${\mc C} = 1$) and a trivial insulator (${\mc C} = 0$), are separated by $1+1$D topological domain walls in spacetime. We use numerical simulations to investigate the universal properties of the DTDMs on these domain walls. 

We first examine the long-time behavior of the DTDMs. Although the dynamics is far from equilibrium, we find that in the long-time limit of each quantum trajectory (labeled by the measurement outcomes), the DTDMs exhibit logarithmic scaling of entanglement and subsystem charge variance, algebraically decaying density correlations, and a definite chirality that resemble those of the ground state of a $1+1$D chiral free-fermion conformal field theory (CFT). Such behavior typically occurs at a domain wall between a Chern insulator and a trivial insulator in an equilibrium Hamiltonian system at zero temperature. The finding that the DTDMs exhibit similar properties in the long-time limit is novel and unexpected, as the adaptive circuit does not explicitly target any particular state at the domain wall. Rather, incompatible measurements generate substantial randomness near the domain wall. This long-time behavior of the DTDMs is controlled by the topologies of the dynamical phases they separate. This behavior of the $1+1$D DTDMs is also nontrivial compared with that of purely $1+1$D charge-conserving free-fermion dynamics, where only area-law entanglement scaling is possible at a finite measurement rate~\cite{PoboikoEtAl2023}. To understand this behavior, we propose a physical picture based on treating the adaptive circuits as cooling algorithms for topological states.

Although the long-time behavior of the DTDMs in individual quantum trajectories resembles that of a chiral free-fermion CFT, their real-time dynamics is quite different. We show that the DTDMs give rise to slow purification dynamics. When the dynamical system starts from a maximally mixed initial state, regions away from the topological domain walls relax to their long-time behavior on an $\mathcal{O}(1)$ circuit timescale, as is typical of area-law-entangled dynamical phases~\cite{gullans2020dynamical}. By contrast, the DTDMs, which experience strongly incompatible measurements, cause the system's entropy to decay slowly and algebraically. We quantify this slow relaxation through the Lyapunov exponents, which govern purification rates. We find that the gap around zero in the Lyapunov spectrum closes with increasing domain-wall length, accounting for the slow purification dynamics.

Finally, we show that the distinctive behavior of the DTDMs is intrinsic to individual quantum trajectories. It is detected by diagnostics that are nonlinear in the trajectory-resolved density matrix. Random measurement outcomes produce different states across trajectories, so directly averaging these states masks the novel behavior of the DTDMs.

The rest of this paper is organized as follows. Section~\ref{sec:Model} defines the circuit and domain-wall geometry. Section~\ref{sec:Results} first validates the dynamical topological phases of the bulk (Sec.~\ref{sec:BulkTopology}). It then presents the entanglement scaling and charge variance of the topological domain walls in the long-time limit and relates them to the spectral density of entanglement energies (Sec.~\ref{sec:PureStates}), resolves the entanglement contribution and chirality of each wall (Sec.~\ref{sec:WallChirality}), and studies the density correlations along the walls and the implied constraints on the charge variance (Sec.~\ref{sec:Correlations}). Next, it examines slow purification and the Lyapunov spectrum (Sec.~\ref{sec:Purification}) and the measurement-outcome-averaged channel dynamics (Sec.~\ref{sec:TrajectoryAveraged}). Finally, Sec.~\ref{sec:picture} interprets the long-time behavior of the walls by viewing the circuit as a local cooling algorithm.
Section~\ref{sec:Conclusions} summarizes the results and discusses open questions. The appendices contain details on trajectory-wise uncertainty analysis and further investigation of finite-size effects.

\section{Topological Adaptive Circuit with Domain Walls}
\label{sec:Model}

We start by reviewing the adaptive circuit construction in Ref.~\cite{BhuiyanPanJian2026} in Sec.~\ref{sec:Stabilizers}, and describe how to incorporate dynamical domain walls into our adaptive circuits.
In Sec.~\ref{sec:Trajectories}, we define the trajectory ensemble and the Gaussian formalism used in our simulations. We refer readers to Ref.~\cite{BhuiyanPanJian2026} for details on the stabilizer formalism and the finite-ancilla adaptive circuit construction. 

% ------------------------------ FIGURE 1 ------------------------------
\begin{figure*}[!t]
    \centering
    \includegraphics[width=0.8\textwidth]{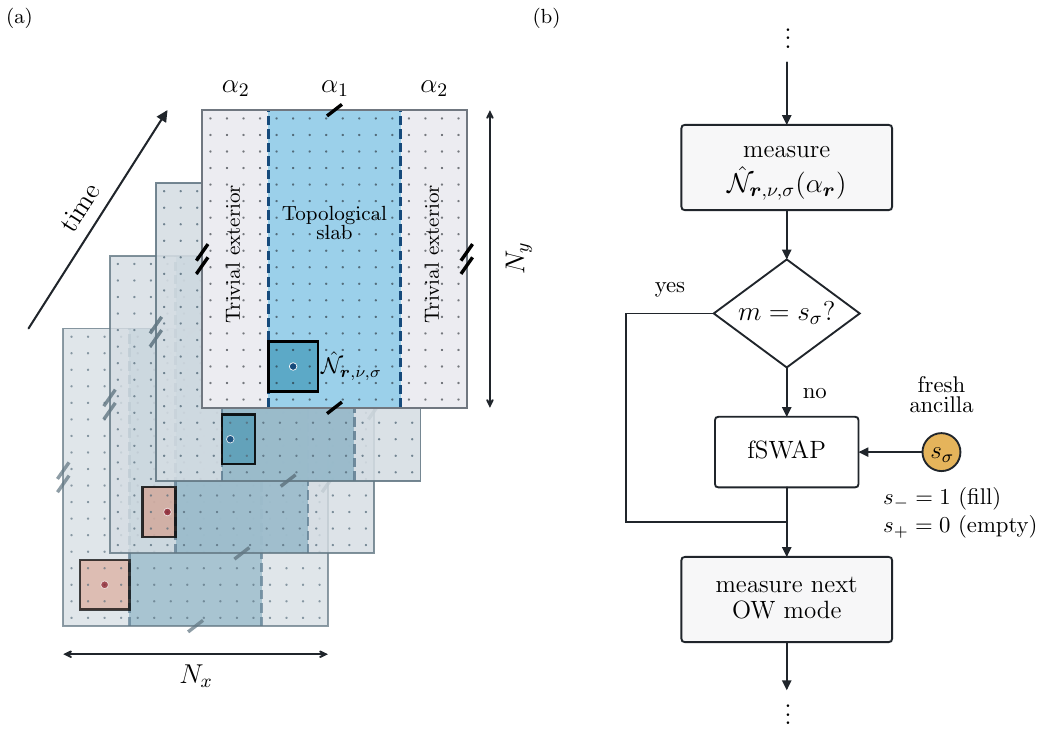}
    \caption{
        \textbf{(a)} Schematic of the domain-wall adaptive circuit and Wannier-mode truncation. Four time slices illustrate measurement updates of OW modes in the periodic geometry, with $\alpha_{\bfr}=\alpha_1$ in the center slab and $\alpha_{\bfr}=\alpha_2$ in the trivial exterior. Dashed lines mark the interfaces, and matching marks identify opposite periodic edges. From the earliest to the latest slice, the highlighted support of the (truncated) OW mode moves from the exterior bulk, across the left interface, into the slab bulk. Red and blue windows show exterior and slab OW supports, respectively, and are clipped at the interface and renormalized. \textbf{(b)} Diagram of a single elementary measurement-feedforward step of the adaptive protocol. An OW-mode occupation measurement $\hat{\mc N}_{\bfr,\nu,\sigma}$ returns the outcome $m=0$ or $1$. If $m$ equals the target occupation $s_\sigma=(1-\sigma)/2$, no correction is applied. Otherwise, an fSWAP with a fresh ancilla sets the mode to $s_\sigma$, which is filled for lower-band modes ($\sigma = -$) and empty for upper-band modes ($\sigma = +$). The procedure then repeats for subsequent OW modes, indicated by the vertical ellipses.
    }
    \label{fig:Geometry}
    \label{fig:Circuit}
\end{figure*}

\subsection{Adaptive circuit construction}
\label{sec:Stabilizers}
\label{sec:Circuit}
\label{sec:FermionCounting}

Based on Ref.~\cite{BhuiyanPanJian2026}, the construction of the adaptive circuit follows from the dynamical bulk-boundary correspondence that relates the topology of the free-fermion dynamics in $2+1$D spacetime to the topology of the long-time states on its temporal boundary, i.e., the final time slice. Here, we focus on charge-conserving dynamics in $2+1$D. Without any additional symmetry, the topology of the dynamics is captured by the Chern number of the long-time states. 

We construct the finite-range topological adaptive circuit that realizes the $2+1$D topological dynamical phases using the following general strategy. We first choose a target Chern-insulator state $\CI$, which can be uniquely specified by a set of occupation constraints on a set of localized modes. Conceptually, the adaptive circuit steers the system toward the target state by repeatedly measuring the occupations of these modes and applying unitary feedforward corrections whenever the outcome is undesired (see Fig.~\ref{fig:Circuit}(b)). Typically, these modes have localized profiles with exponential tails. One needs to truncate them for the construction of the adaptive circuit with only finite-range operations. As a consequence, each trajectory of the circuit is steered toward the target Chern-insulator state $\CI$ with a trajectory-dependent perturbation but the same target Chern number $\mc C$. This Chern number appearing in the long-time limit of the topological adaptive circuit reflects the topology of the spacetime dynamics. Here and henceforth, we will use the double ket $\kett{\cdots}$ to denote many-body states and the single ket $\ket{\cdots}$ to denote single-particle states. 

Here, we take $\CI$ to be the half-filled ground state of a variant of the Qi--Wu--Zhang (QWZ) Hamiltonian~\cite{QiWuZhang2006}, which is defined on a 2D square lattice with two fermion modes $\hat c_{\bfr,\mu=1,2}$ per unit cell $\bfr=(x,y)\in\mathbb Z^2$. With the Fourier convention $\hat c_\mu(\bk)=\sum_{\bfr}e^{\ii\bk\cdot\bfr}\hat c_{\bfr,\mu}$, the QWZ Hamiltonian reads
\begin{align}
    \hat H_{\mr{CI}}&=\int_{\mr{BZ}}\frac{d^2\bk}{(2\pi)^2}\,
    \hat{\bs{c}}^\dagger(\bk)\big[\boldsymbol n(\bk)\cdot\boldsymbol\sigma\big]\hat{\bs{c}}(\bk),\nonumber\\
    \boldsymbol n(\bk)&=(\sin k_x,\ \sin k_y,\ \alpha-\cos k_x-\cos k_y),
    \label{eq:QWZ}
\end{align}
where $\boldsymbol\sigma$ is the vector of Pauli matrices, $\hat{\bs{c}}(\bk)=(\hat c_1(\bk),\hat c_2(\bk))^{\mathsf T}$, and $\alpha$ tunes the target state. The band projectors are $P_\sigma(\bk)=\frac12[\Id+\sigma\,\hat{\boldsymbol n}(\bk)\cdot\boldsymbol\sigma]$, with $\hat{\boldsymbol n}=\boldsymbol n/|\boldsymbol n|$ and band label $\sigma=\pm$. The target state $\CI$ is the half-filled ground state of $\hat H_{\mr{CI}}$, such that the lower band ($\sigma=-$) is completely filled while the upper band ($\sigma=+$) is empty. The Chern number of $\CI$ is $\mc C=1$ for $\alpha\in(0,2)$, $\mc C=-1$ for $\alpha\in(-2,0)$, and $\mc C=0$ for $|\alpha|>2$, with the band gap closing at $\alpha=0,\pm2$. 

A band with nonzero Chern number admits no exponentially localized orthonormal Wannier basis. However, by giving up orthonormality, we can construct an overcomplete basis of exponentially localized Wannier-like modes, which we refer to as overcomplete Wannier (OW) modes~\cite{LiDongLedwithKhalaf2024, BhuiyanPanJian2026}. We therefore describe each band with two exponentially localized OW modes per unit cell, such that we use exactly twice the total number of states. They are obtained by projecting a pair of orthonormal trial orbitals $\tau_\nu$ onto either band,
\begin{equation}
    W_{\bfr,\nu,\sigma}(\bfr',\mu)\propto\int_{\mr{BZ}}\frac{d^2\bk}{(2\pi)^2}\,
    e^{\ii\bk\cdot(\bfr-\bfr')}\big[P_\sigma(\bk)\,\tau_\nu\big]_\mu,
    \label{eq:OWWavefunction}
\end{equation}
where $[\cdots]_\mu$ denotes the $\mu$th component and the proportionality constant normalizes $W_{\bfr,\nu,\sigma}$. We take $\tau_{\mathsf a}=(1,1)^{\mathsf T}/\sqrt2$ and $\tau_{\mathsf b}=(1,-1)^{\mathsf T}/\sqrt2$. The OW mode operators are $\hat\chi^\dagger_{\bfr,\nu,\sigma}=\sum_{\bfr',\mu}W_{\bfr,\nu,\sigma}(\bfr',\mu)\,\hat c^\dagger_{\bfr',\mu}$, and their number operators $\hat{\mc N}_{\bfr,\nu,\sigma}=\hat\chi^\dagger_{\bfr,\nu,\sigma}\hat\chi_{\bfr,\nu,\sigma}$ are projectors. For $\alpha\neq0,\pm2$, $W_{\bfr,\nu,\sigma}$ is exponentially localized around $\bfr$.

The Chern insulator $\CI$ is the unique state satisfying $\hat{\mc N}_{\bfr,\nu,-}\CI=\CI$ and $(\hat\Id-\hat{\mc N}_{\bfr,\nu,+})\CI=\CI$ for all $\bfr$ and $\nu$. These conditions ensure that the lower band of $\hat H_{\mr{CI}}$ is filled and the upper band is empty. We therefore also refer to the number operators as stabilizers. Although the stabilizers of overlapping OW modes do not commute, $\CI$ satisfies all of them simultaneously, so the stabilizing conditions are frustration-free. The adaptive circuit essentially repeatedly measures these stabilizers $\hat{\mc N}_{\bfr,\nu,-}$ and applies subsequent feedforward correction if the measurement outcome does not match the desired outcome for $\CI$. This adaptive circuit dynamics steers the system toward the target state $\CI$ and realizes a topological dynamical phase in $2+1$D. A more formal treatment of these operations will be presented later. 

Now, we introduce the topological dynamical domain walls by letting $\alpha$ depend on the center $\bfr$ of each OW mode. By allowing $\alpha$ to vary in space, adaptive circuits can target states with different Chern numbers in different spatial regions, creating spacetime domain walls between dynamics of different topologies. In this paper, we work on a periodic $N_x\times N_y$ lattice. We choose $\alpha_{\bfr}=\alpha_1$ in a center slab $x_{\text{L}}\leq x\leq x_{\text{R}}$ and $\alpha_{\bfr}=\alpha_2$ in the exterior. We fix $\alpha_2=30$, such that the exterior is deep in the trivial dynamical phase (targeting a trivial insulator with $\mc C =0$), and tune $\alpha_1$. The center slab is bounded by two domain walls of opposite orientation, at $x_{\text{L}}$ and $x_{\text{R}}=x_{\text{L}}+N_x/2$. When $\alpha_1$ is in the range where the target state has $\mc C \neq 0$, the dynamics within the center slab is topological. Consequently, each domain wall sweeps out a $1+1$D worldsheet, separating two topologically distinct dynamical phases. It hosts the nontrivial DTDMs whose behavior will be discussed in Sec. \ref{sec:Results}.

To obtain strictly finite-range operations, we truncate each OW mode about its center and renormalize it:
\begin{equation}
    \widetilde W_{\bfr,\nu,\sigma}(\bfr',\mu)\propto g_{w,\bfr}(\bfr')\,W_{\bfr,\nu,\sigma}(\bfr',\mu),
    \label{eq:TruncatedOW}
\end{equation}
with the choice of window function $g_{w,\bfr}(\bfr')=\mathbf{1}_{\{|x-x'|\leq w\}}\,\mathbf{1}_{\{|y-y'|\leq w\}}\,\mathbf{1}_{\{\alpha_{\bfr}=\alpha_{\bfr'}\}}$. Here, $\mathbf{1}_{X}$ is an indicator function that equals 1 if condition $X$ holds and 0 otherwise, while $\widetilde W_{\bfr,\nu,\sigma}$ denotes the truncated OW modes. The window restricts each mode to a $(2w+1)\times(2w+1)$ square around $\bfr$ and removes its support across any interface [see Fig.~\ref{fig:Geometry}(a) for an illustration]. The domain walls therefore act as open boundaries that separate the center slab and the exterior, even though the lattice remains periodic. Hereafter, $\hat\chi_{\bs{r},\nu,\sigma}$, $\hat{\mc N}_{\bs{r},\nu,\sigma}$, and $W_{\bs{r},\nu,\sigma}$ will denote the fermion operators, number operators, and single-particle wavefunctions of the truncated OW modes.

We remark that, even without domain walls, truncation frustrates the occupation constraints because the truncated modes are no longer built from a single band. As a result, the adaptive circuit dynamics with finite-range operations no longer targets $\CI$ exactly; instead, it steers the system toward an ensemble of states that are essentially $\CI$ with perturbations and share the same Chern number as $\CI$. Despite the slight band mixing and the truncation at the domain wall, we retain the band label $\sigma$ of each truncated OW mode and thus retain its inherited target occupation. 

The adaptive circuits repeatedly measure the occupation of the (truncated) OW modes to compare the system with the target state.
The outcomes of such occupation measurements can be random and need not match these targets. As in Ref.~\cite{BhuiyanPanJian2026} and similar to qubit-based state preparation protocols~\cite{roy2020measurement,puente2024quantum}, the circuit corrects the undesired outcomes with unitary feedforward operations. Each correction applies a fermionic SWAP (fSWAP)~\cite{VerstraeteCiracLatorre2009, CerveraLierta2018} between the measured mode and a fresh fermionic ancillary mode $\hat a$ prepared in the target occupation, which is occupied for a lower-band mode and empty for an upper-band mode---every correction therefore succeeds. In fact, Ref.~\cite{BhuiyanPanJian2026} has constructed this feedforward correction using an alternative ancilla system that shares the same size as the physical dynamical system. The total system conserves charge and thus allows us to apply the topological classification of charge-conserving dynamical systems. However, the feedforward correction only succeeds probabilistically in such a setting. 
Here, for technical simplicity, we introduce a fresh ancillary mode for each measurement to guarantee successful corrections. The topological nature of the dynamics remains the same. Discarding the ancilla reduces each corrective operation to single-fermion gain or loss on the physical system,
\begin{align}
    {}_{a}\braa{0}\fSWAP(\hat\chi_{i,-},\hat a)\big(\hat\Id-\hat{\mc N}_{i,-}\big)\kett{1}_{a}&=\hat\chi^\dagger_{i,-},\nonumber\\
    {}_{a}\braa{1}\fSWAP(\hat\chi_{i,+},\hat a)\,\hat{\mc N}_{i,+}\kett{0}_{a}&=\hat\chi_{i,+},
    \label{eq:GainLoss}
\end{align}
where $i=(\bfr,\nu)$, and $\kett{0}_a$ and $\kett{1}_a$ are the empty and filled states of the ancillary mode.

A measurement of $\hat{\mc N}_{i,\sigma}$ returns $m=1$ with Born probability $\langle\hat{\mc N}_{i,\sigma}\rangle$ and $m=0$ otherwise. By Eq.~\eqref{eq:GainLoss}, each elementary measurement-and-feedforward step [Fig.~\ref{fig:Circuit}(b)] acts on the physical system through the Kraus operators
\begin{align}
    \hat K_{i,-}(m)&=m\,\hat{\mc N}_{i,-}+(1-m)\,\hat\chi^\dagger_{i,-},
    \label{eq:KrausLower}\\
    \hat K_{i,+}(m)&=(1-m)\big(\hat\Id-\hat{\mc N}_{i,+}\big)+m\,\hat\chi_{i,+},
    \label{eq:KrausUpper}
\end{align}
for lower- and upper-band modes, respectively, with $m=0,1$. These satisfy $\sum_{m}\hat K^\dagger_{i,\sigma}(m)\hat K_{i,\sigma}(m)=\hat\Id$, namely, the condition for a positive operator-valued measure, which is required for quantum measurements without postselection~\cite{NielsenChuang2010}. Regardless of the outcome $m$, $\hat K_{i,-}$ leaves the measured lower-band mode filled and $\hat K_{i,+}$ leaves the measured upper-band mode empty.

One complete cycle of our measurement-feedforward protocol visits every unit cell and applies four updates $(\nu,\sigma)\in\{(\mathsf a,+),(\mathsf a,-),(\mathsf b,+),(\mathsf b,-)\}$ per unit cell. For a finite lattice of $N_x\times N_y$ unit cells, one cycle therefore consists of $4N_xN_y$ elementary operations. At fixed truncation range, each cycle can also be implemented in $\mathcal{O}(1)$ circuit depth by grouping the elementary operations into layers of nonoverlapping updates.

\subsection{Quantum trajectories and Gaussian formalism}
\label{sec:Trajectories}

Each run of the circuit produces a quantum trajectory specified by the collection of all the measurement outcomes $\rec$. For an initial state $\hat\rho_0$ and a trajectory $\rec$ accumulated over $t$ cycles, the trajectory-resolved state is $\hat\rho_{\rec}(t)=\hat V_{\rec}\hat\rho_0\hat V_{\rec}^\dagger/p(\rec)$, where $\hat V_{\rec}$ is the ordered product of the Kraus operators along the trajectory and $p(\rec)=\Tr[\hat V_{\rec}\hat\rho_0\hat V_{\rec}^\dagger]$ is the trajectory's Born probability. Because the circuit no longer targets a single state due to the domain wall and the finite-range truncation, $\hat\rho_{\rec}(t)$ only converges toward an ensemble of states at late times. We therefore characterize the stationary ensemble $\{p(\rec),\hat\rho_{\rec}(t)\}$ by trajectory-resolved quantities $O(\hat\rho_{\rec})$ averaged over quantum trajectories,
\begin{equation}
    \overline{O}=\sum_{\rec}p(\rec)\,O(\hat\rho_{\rec}).
    \label{eq:TrajectoryAverage}
\end{equation}
In numerical simulations, the above is approximated by sampling the quantity of interest over a large but finite number of sampled trajectories. Here, $O(\hat\rho_{\rec})$ can be either a linear or nonlinear function of the density matrix $\hat{\rho}_{\rec}$. 

Every Kraus operator in Eqs.~\eqref{eq:KrausLower} and \eqref{eq:KrausUpper} is Gaussian. It creates, annihilates, or projects onto a single mode. Although gain and loss do not conserve particle number, they still map states of definite charge to states of definite charge, even though the charge sector itself changes. Because the dynamics is Gaussian and the state always remains in a definite charge sector, pairing correlations $\langle\hat c_{\mathbf r,\mu}\hat c_{\mathbf r',\mu'}\rangle$ vanish, and the many-body state in each trajectory is therefore fully characterized by its two-point functions
\begin{equation}
    G_{(\bfr,\mu),(\bfr',\mu')}=\Tr\!\big(\hat\rho\,\hat c_{\bfr',\mu'}^\dagger\hat c_{\bfr,\mu}\big).
    \label{eq:TwoPoint}
\end{equation}
When $(\bfr,\mu)$ and $(\bfr',\mu')$ are treated as matrix indices, $G$ is also known as the covariance matrix. The spectrum of $G$ lies in $[0,1]$, and for pure states, $G=G^2$ is the projector onto the occupied single-particle states. 

We write $P_{i,\sigma}=W_{i,\sigma}W_{i,\sigma}^\dagger$ for the projector onto the truncated OW mode with index $(i,\sigma)$. The update of the physical covariance matrix following an occupation measurement of mode $(i,\sigma)$ with outcome $m\in\{0,1\}$ and the outcome-conditioned feedforward correction is
\begin{align}
    G'={}&(\Id-P_{i,\sigma})G(\Id-P_{i,\sigma})+s_\sigma P_{i,\sigma}\nonumber\\
    &-\frac{2m-1}{p_{i,\sigma}(m)}(\Id-P_{i,\sigma})GP_{i,\sigma}G(\Id-P_{i,\sigma}),
    \label{eq:CovarianceUpdate}
\end{align}
where the target occupation is $s_\sigma=(1-\sigma)/2\in\{0,1\}$, and $p_{i,\sigma}(m)=1-m+(2m-1)\operatorname{tr}(P_{i,\sigma}G)$ is the Born probability for outcome $m\in\{0,1\}$~\cite{bravyi2004lagrangian}. The adaptive circuit is then simulated by iterating Eq.~\eqref{eq:CovarianceUpdate}, with an entire cycle visiting all modes $(i,\sigma)=(\bfr,\nu,\sigma)$.

% ------------------------------ SECTION III ---------------------------
\section{Chiral Dynamical Topological Domain-Wall Modes}
\label{sec:Results}

First, we confirm that the center slab hosts a topological dynamical phase that targets a Chern insulator with Chern number $\mc C =1$ (Sec.~\ref{sec:BulkTopology}). We then show that in the long-time limit, the DTDMs on each topological domain wall carry entanglement, charge variance, chirality, and spatial correlations that resemble those of the ground state of a single chiral free fermion in $1+1$D (Secs.~\ref{sec:PureStates}, \ref{sec:WallChirality}, \ref{sec:Correlations}). Next, we show that the DTDMs experience slow purification dynamics quantified by
a small Lyapunov gap that closes as the inverse of the length of the domain wall $N_y^{-1}$ (Sec.~\ref{sec:Purification}). Finally, we show that the novel features of the DTDMs are not visible in the trajectory-averaged state (Sec.~\ref{sec:TrajectoryAveraged}).
We provide a physical picture of the long-time behavior of the DTDMs based on treating the adaptive dynamics as a cooling algorithm in Sec.~\ref{sec:picture}.

In the following calculation, we simulate the circuit dynamics numerically with truncation range $w=1$. All quantities, linear or nonlinear in the density matrix, are averaged over 100 trajectories unless stated otherwise.

\subsection{Bulk topology}
\label{sec:BulkTopology}

\begin{figure}[!t]
    \StackedFigure
    \centering
    \includegraphics[width=\columnwidth]{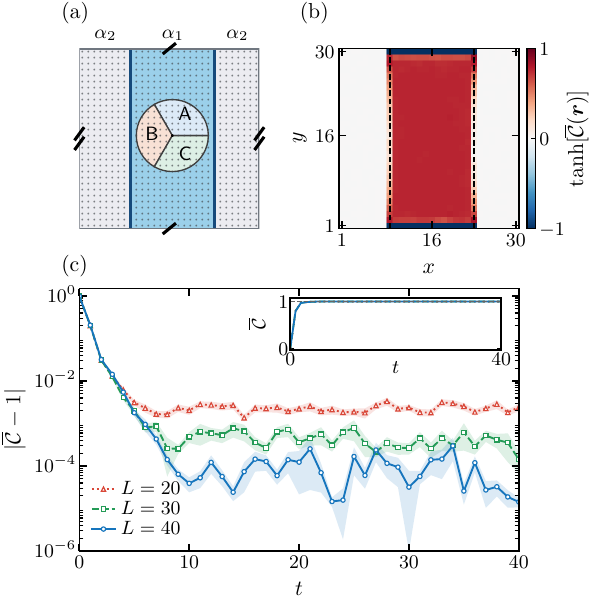}
    \caption{\textbf{(a)} Geometry of the real-space Chern number. A $30\times 30$ lattice with the center slab (blue) and the tripartite disk. Slashes mark periodic identifications. \textbf{(b)} Local Chern marker $\overline{\mathcal C}(\bfr)$ at $t=40$, normalized by $\tanh$ to compress the boundary artifact of the position operators on the periodic lattice. \textbf{(c)} Convergence of the real-space Chern number deviation from unity, $|\overline{\mc{C}}-1|$ for $L=20,30,40$, with $\overline{\mc{C}}$ in the inset. Error bands here and elsewhere (including error bars) on trajectory-averaged quantities denote the standard error of the mean over sampled trajectories.}
    \label{fig:BulkTopology}
\end{figure}

We start by confirming that the adaptive circuit realizes the topological dynamical phase that steers the system toward a Chern insulator in the center slab. Let $\ket{\bfr,\mu}$ denote the single-particle state localized on orbital $\mu$ of unit cell $\bfr$. We divide a disk inside the slab into three regions $\mathsf A,\mathsf B,\mathsf C$ as shown in Fig.~\ref{fig:BulkTopology}(a). $\Pi_{\mathsf S}=\sum_{\bfr\in\mathsf S}\sum_\mu\ketbra{\bfr,\mu}$ denotes the projector onto a given set of unit cells $\mathsf S$. For each trajectory, the real-space Chern number formula is given by~\cite{Kitaev2006}
\begin{equation}
    \mc C=-24\pi\operatorname{Im}\operatorname{tr}\!\big(\Pi_{\mathsf A} G\,\Pi_{\mathsf B} G\,\Pi_{\mathsf C} G\big).
    \label{eq:KitaevChern}
\end{equation}
For $\mc C$ to be quantized, the disk should lie inside the slab, stay away from the interfaces, and be much larger than the system's correlation length within the slab. We therefore operationally take the disk radius $R=0.2L$ at $\alpha_1=1$ in square
systems with $N_x = N_y =L$. Starting from random half-filled states, we run our adaptive circuit for $40$ cycles for each system size $L$, finding that $\overline{\mc C}$ converges to unity within about ten cycles, with a residual deviation that decreases with $L$ [Fig.~\ref{fig:BulkTopology}(c)].

In addition, we compute the local Chern marker, which spatially resolves the topology~\cite{BiancoResta2011}
\begin{equation}
\mathcal{C}(\bfr)=-2\pi\sum_{\mu}\mathrm{Im}\expval{\comm{GXG}{GYG}}{\bfr,\mu},
 \label{eq:LocalMarker}
\end{equation}
with $X=\sum_{\bfr,\mu}x\ketbra{\bfr,\mu}$ and $Y=\sum_{\bfr,\mu}y\ketbra{\bfr,\mu}$ the position operators in each direction. We confirm that our adaptive circuit steers the system toward steady states with a local Chern marker that is uniform across the slab and vanishes in the exterior, revealing two dynamical topological domain walls [Fig.~\ref{fig:BulkTopology}(b)].

\begin{figure*}[!t]
    \StackedFigure
    \centering
    \includegraphics[width=\textwidth]{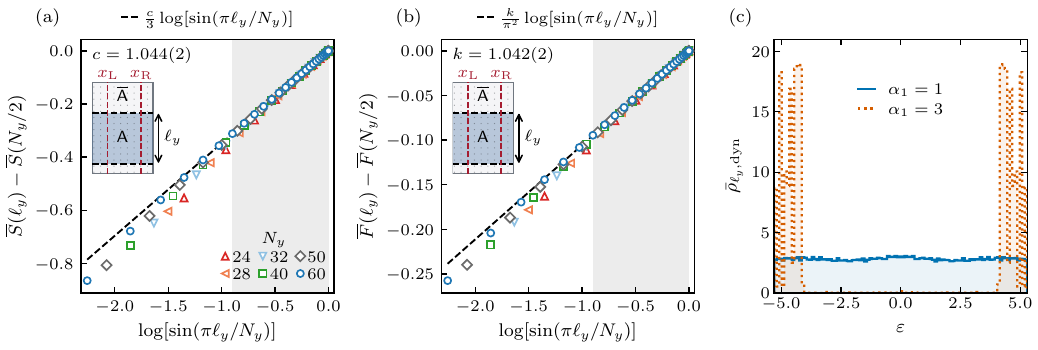}
    \caption{\textbf{(a)} Strip entanglement entropy relative to its half-strip value for $\alpha_1=1$ and several $N_y$ values. The inset shows the strip $\mathsf{A}$ crossing both walls (red dashed lines), with $\overline{\mathsf{A}}$ the complement. The dashed black line is a joint fit to Eq.~\eqref{eq:CFTScaling}, giving $c=1.044(2)$. \textbf{(b)} Same analysis for the charge variance, giving $k=1.042(2)$. \textbf{(c)} Average spectral density $\bar{\rho}_{\ell_y,\mathrm{dyn}}(\varepsilon)$ of half-strip ($\ell_y=N_y/2$) entanglement energies per subsystem for $\alpha_1=1$ and $\alpha_1=3$ at $N_y=32$, pooled over 100 trajectories and all 32 subsystem origins. Only entanglement energies with $|2\nu(\varepsilon)-1|<0.99$ are shown.} 
    \label{fig:EntropyCharge}
\end{figure*}

\subsection{Entanglement and charge variance}
\label{sec:PureStates}
\label{sec:Entanglement}

We use the entanglement entropy and the charge variance of strip subsystems to probe the long-time behavior of the DTDMs hosted by the domain walls (see inset of Fig.~\ref{fig:EntropyCharge}(a)). We find that both grow logarithmically with the strip length, with coefficients matching those of the ground state of a $1+1$D free-fermion CFT, and we trace this agreement to the spectral density of entanglement energies.

As shown in Sec.~\ref{sec:BulkTopology}, the slab bulk and the exterior are both short-range correlated. We therefore consider strip subsystems $\mathsf{A}$ of length $\ell_y$ running parallel to the domain walls that span the full width of the system $\mathsf{A}(y_0,\ell_y)=\{(x,y):(y-y_0)\bmod N_y<\ell_y\}$, such that each strip crosses both walls [Fig.~\ref{fig:EntropyCharge}(a) inset]. The strip $\mathsf{A}$ contains an interval of length $\ell_y$ along each domain wall, which we will vary in our calculations. We keep the transverse width $N_x$ fixed throughout. With such a choice of $\mathsf{A}$, our subsequent characterizations treat the 2D lattice as a quasi-1D system along the $y$ direction. 

For any subsystem $\mathsf{A}$, we denote the reduced density matrix by $\hat\rho_{\mathsf{A}}$ and the covariance matrix restricted to $\mathsf{A}$ by $G_{\mathsf{A}}$. The entanglement entropy of $\mathsf{A}$ and the variance of the subsystem charge $\hat N_{\mathsf{A}}$ are given by $S(\mathsf{A})=-\Tr\hat\rho_{\mathsf{A}}\log\hat\rho_{\mathsf{A}}$ and $F(\mathsf{A})=\langle\hat N_{\mathsf{A}}^2\rangle-\langle\hat N_{\mathsf{A}}\rangle^2$, respectively. For a Gaussian state, $\hat\rho_{\mathsf{A}}$ is fully determined by $G_{\mathsf{A}}$. $S(\mathsf{A})$ and $F(\mathsf{A})$ reduce to
\begin{align}
S(\mathsf{A})&=-\operatorname{tr}\big[G_{\mathsf{A}}\log G_{\mathsf{A}}+(\mathbf{1}-G_{\mathsf{A}})\log(\mathbf{1}-G_{\mathsf{A}})\big],\label{eq:entropy}\\
F(\mathsf{A})&=\operatorname{tr}\big[G_{\mathsf{A}}(\mathbf{1}-G_{\mathsf{A}})\big].\label{eq:chargevariance}
\end{align}
We will numerically compute $S(\mathsf{A})$ and $F(\mathsf{A})$ averaged over the sampled quantum trajectories and over all translated subsystem origins~\footnote{As a numerical trick, averaging over $y_0$ within each trajectory speeds up the convergence of the simulation. We apply similar averaging throughout the paper.}, and will denote the averaged quantities as $\overline{S}(\ell_y)$ and $\overline{F}(\ell_y)$. This notation is chosen to emphasize their dependence on the length $\ell_y$ of the domain-wall interval included in $\mathsf{A}$. 

We find that both the entanglement entropy and the charge variance grow logarithmically with the strip length $\ell_y$, which is similar to the behavior of ground states in critical Hamiltonian systems in $1+1$D. Due to this similarity, we numerically fit this growth using the following entanglement entropy and subsystem charge variance formulae motivated by $1+1$D critical equilibrium ground states on a ring
\begin{align}
    \overline{S}(\ell_y)&=\frac{c}{3}\log D(\ell_y)+\text{const.},\nonumber\\
    \overline{F}(\ell_y)&=\frac{k}{\pi^2}\log D(\ell_y)+\text{const.},
    \label{eq:CFTScaling}
\end{align}
where $k$ and $c$ are treated as fitting parameters. Here, $D(\ell_y)=(N_y/\pi)\sin(\pi\ell_y/N_y)$ is the chord length, which accounts for periodicity along the length of the domain wall~\cite{CalabreseCardy2004}, and the constant terms encode non-universal information and bulk area-law contributions (which are subleading in $\ell_y$ and can depend on $N_x$). Note that if we were to fit these parameters $c$ and $k$ for $1+1$D critical ground states in equilibrium, we would obtain the central charge and the level of the $U(1)$ Kac--Moody current algebra of the underlying CFT~\cite{CalabreseCardy2004,song2012bipartite,DiFrancescoMathieuSenechal1997}. Eliminating the constant terms in Eq.~\eqref{eq:CFTScaling} by subtracting each trajectory's half-strip value at $\ell_y=N_y/2$, the subtracted curves for all $N_y$ collapse onto straight lines in $\log\sin(\pi\ell_y/N_y)$ [Fig.~\ref{fig:EntropyCharge}(a,b)], giving $c=1.044(2)$ and $k=1.042(2)$. Since the slab bulk and the exterior are short-range correlated, the entanglement entropy and charge variance in Eq.~\eqref{eq:CFTScaling} should be attributed to nontrivial modes, i.e., the DTDMs, living on the $1+1$D dynamical topological domain walls in our system.

To put these fitting parameters $c$ and $k$ in perspective, consider a conceptually related equilibrium Hamiltonian system where a Chern insulator with Chern number $\mc C=1$ occupies the center slab region and a trivial insulator occupies the exterior. Each domain wall between the Chern and trivial insulators hosts a chiral free-fermion mode. When the whole system is treated as a quasi-1D system, the universal properties of its ground state are essentially equivalent to those of a $1+1$D gapless free-fermion chain carrying a pair of counterpropagating fermion modes, and it can be described by a $1+1$D free-fermion CFT. Calculating the entanglement entropy $S(\mathsf{A})$ and the charge variance $F(\mathsf{A})$ for the same strip subsystem $\mathsf{A}$ in this equilibrium system would yield the same behavior shown in Eq.~\eqref{eq:CFTScaling} with parameters $c=1$ and $k=1$.  As mentioned, for a ground state of an equilibrium CFT, these parameters are the central charge and the level of the current algebra. They are universal parameters that are not continuously tunable in equilibrium.

Hence, the long-time behavior of the DTDMs in our dynamical systems strikingly resembles that of the free-fermion CFT ground state in terms of entanglement-entropy scaling and charge variances. Bear in mind that the former results from nontrivial dynamics with topological domain walls in spacetime. Its long-time behavior is independent of the initial state but is subject to randomness induced by the measurements. In contrast, the CFT ground state is the lowest-energy eigenstate of a critical static Hamiltonian. \textit{A priori}, there is no obvious reason the dynamical system should produce entanglement and charge variances similar to those of an equilibrium system. Hence, the scaling behavior in Eq.~\eqref{eq:CFTScaling} with parameters $c$ and $k$ identical to those of an equilibrium CFT is a nontrivial feature of the long-time behavior of the DTDMs. (The small excess of the fitted $c$ and $k$ over one for the DTDMs is likely a finite-size effect, which we examine further in App.~\ref{app:SupplementaryNumerics}.)

Another interesting perspective comes from comparing our dynamical system, treated as a quasi-1D system, with a purely $1+1$D free-fermion dynamical system with charge conservation. As shown in Ref.~\cite{PoboikoEtAl2023}, the latter systems exhibit area-law entanglement entropy scaling at long times for any nonzero measurement rate. In contrast, the $1+1$D DTDMs in our system can exhibit logarithmic entanglement scaling in the long-time limit, which we attribute to the topological nature of each domain, i.e., each domain wall separates two topologically distinct $2+1$D dynamical phases.

In fact, the resemblance between the long-time behaviors of the DTDMs and the equilibrium free-fermion CFT ground state goes beyond the logarithmic scaling behaviors in Eq.~\eqref{eq:CFTScaling}. For a Gaussian state, the entanglement entropy $S(\mathsf{A})$ and charge variance $F(\mathsf{A})$ are both completely determined by the distribution of the eigenvalues $\nu_a$ of the covariance matrix $G_{\mathsf{A}}$ (with $a$ indexing the eigenvalues). Each eigenvalue $\nu_a$ can be re-expressed through its corresponding single-particle entanglement energy $\varepsilon_a$ as
\begin{equation}
    \nu_a =\frac{1}{1+e^{\varepsilon_a}},
    \label{eq:EntanglementEnergies}
\end{equation}
whose spectral density is defined as 
\begin{align}
\rho_{\mathsf{A}}(\varepsilon) \equiv \sum_a \delta(\varepsilon - \varepsilon_a).    
\end{align}
Next, we compare the long-time behaviors of the DTDMs and the ground state of the free-fermion CFT using this spectral density.

We use $\bar{\rho}_{\ell_y,\rm dyn}$ to denote $\rho_{\mathsf{A}}$ for the specified strip subsystem in the long-time limit of our circuit dynamics, averaged over quantum trajectories and translated subsystem origins. The subscript ``dyn" emphasizes that this spectral density $\bar{\rho}_{\ell_y,\rm dyn}$ is obtained from our adaptive dynamics. The averaged entanglement entropy $\overline{S}(\ell_y)$ and the charge variance $\overline{F}(\ell_y)$ can be rewritten as 
\begin{align}
     & \overline{S}(\ell_y) = \int  d\varepsilon~ \bar{\rho}_{\ell_y,\rm dyn}(\varepsilon) \left[ \log (1+e^\varepsilon) - \frac{\varepsilon e^\varepsilon}{1+e^\varepsilon}\right], \nonumber \\
     & \overline{F}(\ell_y) = \int  d\varepsilon~ \bar{\rho}_{\ell_y,\rm dyn}(\varepsilon) \frac{e^\varepsilon}{(1+e^\varepsilon)^2}.
     \label{eq:SF_integral}
\end{align}
Numerically, we find that $\bar{\rho}_{\ell_y,\rm dyn} (\varepsilon)$ is well approximated by the ansatz
\begin{align}
   \bar{\rho}_{\ell_y,\rm dyn} (\varepsilon) = \frac{\zeta}{\pi^2} \log D(\ell_y) +  \text{const.} \label{eq:rho_ansatz}
\end{align}
in a large window $\varepsilon \in [-\Lambda, \Lambda]$ of the entanglement energy. The ``const.'' part is subleading in $\ell_y$ and independent of $\varepsilon$. Here, $\zeta$ is a constant (independent of $\varepsilon$ and $\ell_y$). An example of $\bar{\rho}_{\ell_y,\rm dyn}$ with $\alpha_1 = 1$ (where the dynamics targets a Chern insulator in the slab bulk) is shown in Fig.~\ref{fig:EntropyCharge}(c). (In the same figure, as a control, the constant behavior of $\bar{\rho}_{\ell_y,\rm dyn}$ disappears for $\alpha_1 = 3$, i.e., when the dynamics targets a trivial insulator in the center slab.) This ansatz naturally leads to logarithmic scaling of $ \overline{S}(\ell_y)$ and $ \overline{F}(\ell_y)$ [Eq.~\eqref{eq:CFTScaling}] and the relations
\begin{align}
    & c =  \frac{3\zeta}{\pi^2} \int_{-\Lambda}^{\Lambda} d\varepsilon~   \left[ \log (1+e^\varepsilon) - \frac{\varepsilon e^\varepsilon}{1+e^\varepsilon}\right] = \zeta + O(e^{-\Lambda}), \nonumber \\
    & k = \zeta \int_{-\Lambda}^{\Lambda} d\varepsilon~\frac{e^\varepsilon}{(1+e^\varepsilon)^2} =  \zeta + O(e^{-\Lambda}).
\end{align}
Technically, we have assumed that the behavior of $\bar{\rho}_{\ell_y,\rm dyn} (\varepsilon)$ beyond the window $[-\Lambda, \Lambda]$ does not overcome the exponential decay $\sim e^{-|\varepsilon|}$ in the rest of the integrand in Eq.~\eqref{eq:SF_integral}. The earlier fitted parameters $c=1.044(2)$ and $k=1.042(2)$ indicate that $\zeta$ is very close to $1$ for the long-time behavior of our dynamical system with domain walls.

The ansatz Eq.~\eqref{eq:rho_ansatz} is, in fact, inspired by results in the $1+1$D free-fermion CFT~\cite{EislerPeschel2013, CalabreseCardy2004}. For the ground state of the free-fermion CFT on a ring of length $N_y$, the entanglement energy spectral density $\rho_{\ell_y,\rm CFT}$ associated with an interval of length $\ell_y$ is given by the constant function $\rho_{\ell_y,\rm CFT} (\varepsilon) = \frac{1}{\pi^2} \log D(\ell_y) + \text{const.}$ In other words, $\zeta = 1$ is an exact result for the free-fermion CFT ground state.

If $\bar{\rho}_{\ell_y,\rm dyn} $ indeed follows Eq.~\eqref{eq:rho_ansatz} with exactly $\zeta=1$ for our dynamical system with domain walls, the behavior of any scalar function $f(G_{\mathsf{A}})$ of the covariance matrix $G_{\mathsf{A}}$ in the long-time limit of our circuit dynamics will be identical to that in the ground state of the free-fermion CFT (assuming $f$ decays exponentially outside the window $\varepsilon \in [-\Lambda, \Lambda]$). The entanglement entropy and the charge variance provide two examples of such scalar functions.

\subsection{Chirality of the dynamical topological domain walls}
\label{sec:WallChirality}

\begin{figure}[!t]
    \centering
    \includegraphics[width=3.16in]{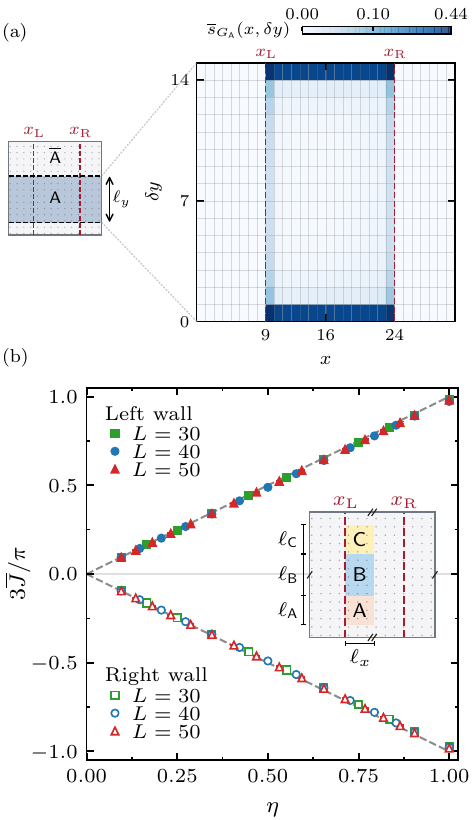}
    \caption{\textbf{(a)} Trajectory-averaged entanglement entropy contour $\overline{s}_{G_{\mathsf{A}}}(x,\delta y)$ of the half strip $\mathsf{A}$ (left) for $L=30$. Red dashed lines mark the walls. Here $\delta y=(y-y_0)\bmod N_y$ is the displacement along $y$ from the starting row $y_0$ of the strip. We used a square-root colorbar to emphasize the entanglement contour along the domain wall. \textbf{(b)} Rescaled trajectory-averaged modular commutator $3\overline J/\pi$ versus the cross ratio $\eta$ at the left (filled markers) and right (open markers) walls, for regions $\mathsf{A}$, $\mathsf{B}$, $\mathsf{C}$ along one wall (inset). Gray dashed lines mark $3\overline J/\pi=\pm\eta$.} 
    \label{fig:ChiralModular}
\end{figure}

The strip subsystem $\mathsf{A}$ considered in Sec.~\ref{sec:Entanglement} 
crosses both domain walls, so the entanglement entropy and charge variance capture only their combined contribution. Both quantities are also even under the change of spatial orientation (and time reversal) and therefore blind to chirality. We now resolve the contribution of each wall with the entanglement contour, and probe its chirality directly with the modular commutator~\cite{KimShiKatoAlbert2022,KimShiKatoAlbert2022b,ZouShiSorceLimKim2022}. The modular commutator is sensitive to the spatial orientation and can extract the chiral central charge from the bulk of a gapped $2+1$D chiral state and its edge. In the subsequent discussions, we set $\alpha_1 = 1$ for the center slab.

To examine the individual entanglement contributions of the DTDMs along each domain wall, we use the entanglement entropy contour, which spatially resolves the entanglement entropy of a chosen subsystem $\mathsf{A}$~\cite{ChenVidal2014}. For a Gaussian state with subsystem covariance matrix $G_{\mathsf{A}}$, the entanglement entropy contour is defined as
\begin{equation}
    s_{G_{\mathsf{A}}}(\bfr)=-\sum_\mu\big[G_{\mathsf{A}}\log G_{\mathsf{A}}+(\mathbf{1}-G_{\mathsf{A}})\log(\mathbf{1}-G_{\mathsf{A}})\big]_{(\bfr,\mu),(\bfr,\mu)},
    \label{eq:Contour}
\end{equation}
such that $s_{G_{\mathsf{A}}}(\bfr)$ is nonnegative and satisfies $\sum_{\bfr\in\mathsf{A}}s_{G_{\mathsf{A}}}(\bfr)=S(\mathsf{A})$, with $S(\mathsf{A})$ being the entanglement entropy [Eq.~\eqref{eq:entropy}]. Figure~\ref{fig:ChiralModular}(a) shows the trajectory-averaged entropy contour $\overline{s}_{G_{\mathsf{A}}}$ for the strip subsystem of Sec.~\ref{sec:Entanglement}, with $\ell_y=N_y/2$. Apart from the area-law contribution along the cuts, the contour is concentrated along the two domain walls with equal weight and extends along the length of the domain wall. Because of our specific choice of the OW modes truncation at the domain wall, the center slab and the exterior decouple after a single cycle of the adaptive circuit. The DTDM contributions therefore lie entirely on the slab side of each wall. We remark that these microscopic details of the domain wall do not affect the universal behavior discussed below. With $\alpha_1 =1$, summing over $s_{G_{\mathsf{A}}}(\bfr)$ within a neighborhood of a domain wall segment in $\mathsf{A}$ shows that the DTDMs along each domain wall carry half of the logarithmic entanglement examined in Fig.~\ref{fig:EntropyCharge}(a), i.e., $\frac16\log D(\ell_y)$, consistent with the ground state of a single chiral (or anti-chiral) sector of a $1+1$D free-fermion CFT~\cite{HolzheyLarsenWilczek1994, CalabreseCardy2004}.

We further test this claim and probe the chirality of the $1+1$D DTDMs directly via the modular commutator
\begin{equation}
    J(\mathsf{A},\mathsf{B},\mathsf{C})=\ii\Tr\big(\hat\rho_{\mathsf{ABC}}\big[\hat H_{\mathsf{AB}},\hat H_{\mathsf{BC}}\big]\big),
    \label{eq:ManyBodyModularCommutator}
\end{equation}
where $\hat H_{\mathsf{S}}=-\log\hat\rho_{\mathsf{S}}$ is the modular Hamiltonian of the reduced density matrix $\hat\rho_{\mathsf{S}}$ of region $\mathsf{S}$, and $\mathsf{AB}=\mathsf{A}\cup\mathsf{B}$, with $\mathsf{BC}$ and $\mathsf{ABC}$ defined similarly. For a Gaussian state, Eq.~\eqref{eq:ManyBodyModularCommutator} reduces to~\cite{FanZhangGu2023,ZouShiSorceLimKim2022}
\begin{equation}
J(\mathsf{A},\mathsf{B},\mathsf{C})=\ii\operatorname{tr}\big(G_{\mathsf{ABC}}\big[h_{\mathsf{AB}},h_{\mathsf{BC}}\big]\big),
    \label{eq:GaussianModularCommutator}
\end{equation}
where the single-particle modular Hamiltonians are determined by the corresponding subsystem covariance matrices through $h_{\mathsf{S}}=\log\big[(\mathbf{1}-G_{\mathsf{S}})G^{-1}_{\mathsf{S}}\big]$~\footnote{In Eq.~\eqref{eq:GaussianModularCommutator}, the single-particle modular Hamiltonians $h_{\mathsf{AB}}$ and $h_{\mathsf{BC}}$ are padded by zeros to the modes of $\mathsf{ABC}$.}.

As shown in Ref.~\cite{ZouShiSorceLimKim2022}, the modular commutator of three contiguous regions $\mathsf{A}$, $\mathsf{B}$, $\mathsf{C}$ along the edge of a $2+1$D chiral topological ground state of a gapped Hamiltonian system takes the universal form
\begin{equation}
    J(\mathsf{A},\mathsf{B},\mathsf{C})=\frac{\pi c_-}{3}\eta, \quad \eta=\frac{D(\ell_{\mathsf{A}})\,D(\ell_{\mathsf{C}})}{D(\ell_{\mathsf{AB}})\,D(\ell_{\mathsf{BC}})},
    \label{eq:RingCrossRatio}
\end{equation}
where $\eta$ is the cross ratio, $\ell_{\mathsf{S}}$ is the length of region $\mathsf{S}$ along $y$, and $c_-$ is the chiral central charge. Equation~\eqref{eq:RingCrossRatio} assumes that each region extends into the bulk with a width $\ell_x$ much greater than the bulk correlation length. Furthermore, the sign of $c_-$ is sensitive to the ordering of $\mathsf{A}$, $\mathsf{B}$, $\mathsf{C}$ along the interface. Reversing the ordering changes the sign of $c_-$. Here, we order the regions along the $+y$ direction on both walls [inset of Fig.~\ref{fig:ChiralModular}(b)]. Thus, if the DTDMs along the two walls have opposite chiralities, the modular commutator should give opposite signs of $c_-$ for the two walls.

For the results in Fig.~\ref{fig:ChiralModular}, we switch from the $N_x=20$ systems of Sec.~\ref{sec:Entanglement} to square systems with $N_x=N_y=L$, so that the slab widens with the wall length~\footnote{Unlike the entanglement entropy and charge variance, the modular commutator turns out to be more susceptible to finite-size effects in both the wall length \textit{and} the wall separation.}. We evaluate Eq.~\eqref{eq:GaussianModularCommutator} at $t=2L$ and take each region to extend a width $\ell_x=\lfloor L/4\rfloor$ from the wall into the slab. As discussed above, the DTDMs lie entirely on the slab side of each wall [Fig.~\ref{fig:ChiralModular}(a)], so the regions need not extend into the exterior.  To examine the dependence of $J$ on the cross-ratio $\eta$, we set $\ell_{\mathsf{A}}=\ell_{\mathsf{C}}$ and $\ell_{\mathsf{B}}=N_y/2-\ell_{\mathsf{A}}$, such that $\eta=\sin^2(\pi\ell_{\mathsf{A}}/N_y)$ can be varied by changing $\ell_{\mathsf{A}}$. To compute $J$ for $\eta=1$, we divide the entire wall into three nearly equal contiguous intervals. 
For each trajectory, we average $J$ over all $N_y$ translations of the partitions along the wall, and then average over trajectories to obtain $\overline J$.

Figure~\ref{fig:ChiralModular}(b) shows that $3\overline J/\pi$ follows $+\eta$ at the left wall and $-\eta$ at the right wall for all $L$. The two domain walls therefore host $1+1$D chiral DTDMs of opposite chirality $c_{-}=\pm1$. 
The boundary values $c_-=\pm1$ thus match the jumps in Chern number across the two walls, confirming the bulk-boundary correspondence directly within each quantum trajectory and without reference to a parent Hamiltonian.

\subsection{Correlations along the domain walls}
\label{sec:Correlations}

\begin{figure*}[!t]
    \StackedFigure
    \centering
    \includegraphics[width=\textwidth]{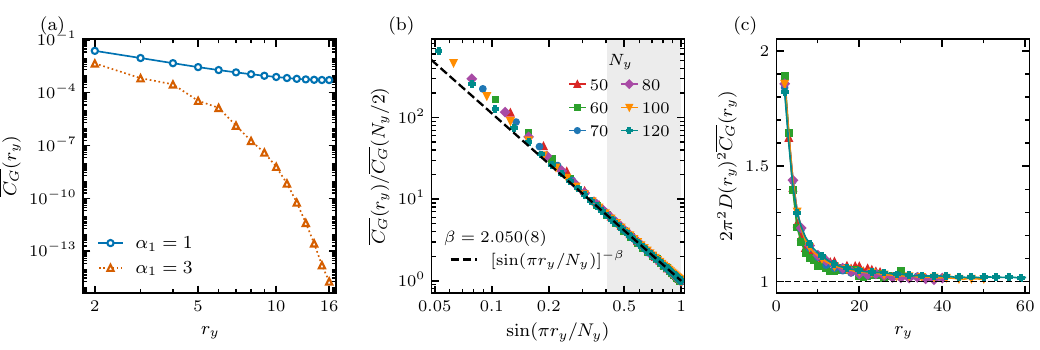}
    \caption{\textbf{(a)} Squared correlator at $N_y=32$ for $\alpha_1=1$ and $3$. \textbf{(b)} Squared correlator normalized by its antipodal value for $\alpha_1=1$ and several $N_y$. The dashed line is a joint fit to $[\sin(\pi r_y/N_y)]^{-\beta}$, giving $\beta=2.050(8)$. \textbf{(c)} Rescaled squared correlator $2\pi^2D(r_y)^2\,\overline{C_G}(r_y)$ versus separation $r_y$ for the same sizes and at $\alpha_1=1$, with $D(r_y)$ the chord length. The dashed line marks the two-wall prediction, Eq.~\eqref{eq:TwoWallPrediction}, with $k=1$.}
    \label{fig:Correlations}
\end{figure*}

In the following, we show that the density correlation along the dynamical topological domain walls exhibits a power-law behavior that corroborates the logarithmic scaling of the charge variance discussed in Sec.~\ref{sec:Entanglement}. As in Sec.~\ref{sec:Entanglement}, we fix $N_x=20$ and evolve for $t=2N_y$ cycles.

Within each quantum trajectory, the equal-time connected density correlation in the Gaussian state is $\langle\hat n_{\bfr,\mu}\hat n_{\bfr',\mu'}\rangle_c=-|G_{(\bfr,\mu),(\bfr',\mu')}|^2$ for $(\bfr,\mu) \neq (\bfr',\mu')$. To probe the density correlations contributed by the domain walls, we define the squared correlator
\begin{align}
    C_G(r_y)&=\frac{1}{N_y}\sum_{y,x,x'}\sum_{\mu,\mu'}\big|G_{(x,y,\mu),(x',y+r_y,\mu')}\big|^2\nonumber\\
    &=-\frac{1}{N_y}\sum_y\langle\hat n_y\hat n_{y+r_y}\rangle_c,
    \label{eq:SquaredCorrelator}
\end{align}
where $\hat n_y=\sum_{x,\mu}\hat n_{x,y,\mu}$, and where the second equality holds for $r_y\neq0$ (mod $N_y$)~\footnote{The restriction $r_y\neq0$ (mod $N_y$) for the equivalence of the squared correlator with minus the connected density correlator follows from Wick's theorem.}. Furthermore, summing over $x$ and $x'$ captures the total density of the DTDMs, which extend over a finite width around each wall.

We now study the long-time behavior of $C_G$ averaged over trajectories, which we denote by $\overline{C_G}$. Figure~\ref{fig:Correlations}(a) shows that $\overline{C_G}$ appears to decay algebraically for $\alpha_1=1$, and decays exponentially for the control $\alpha_1=3$, implying that the long-range correlations must be contributions of the DTDMs at the topological domain walls.

We quantify these density correlations further with a single-parameter power-law fit. Motivated by the similarity between the long-time behavior of DTDMs and the free-fermion CFT ground state shown in Sec.~\ref{sec:Entanglement}, we expect the former to exhibit the power-law behavior $\overline{C_G}\propto D(r_y)^{-2}$, which is exactly the behavior of the equal-time density correlation in a free-fermion CFT on a ring. To test this numerically, we normalize $\overline{C_G}$ by its value at the antipodal separation $r_y=N_y/2$ and fit it to $[\sin(\pi r_y/N_y)]^{-\beta}$ for all $N_y$ [Fig.~\ref{fig:Correlations}(b)]. This fitting yields $\beta=2.050(8)$, which is very close to the expected $\beta=2$ value. Next, we show that the exponent $\beta = 2$ is crucial for the logarithmic scaling of the charge variance shown in Sec.~\ref{sec:Entanglement}.

To connect the charge variance and $\overline{C_G}$, we first rewrite the charge variance $F(\mathsf{A})$ in a generic subsystem $\mathsf{A}$ for a pure Gaussian state:
\begin{equation}
    F(\mathsf{A})=\sum_{(\bfr,\mu)\in \mathsf{A}}~\sum_{(\bfr',\mu')\notin \mathsf{A}}|G_{(\bfr,\mu),(\bfr',\mu')}|^2. 
    \label{eq:chargevariancev2}
\end{equation}
This rewriting uses the definition in Eq.~\eqref{eq:chargevariance} and the fact that $G^2 = G$ for a pure Gaussian state. Applying this equation to the strip subsystem $\mathsf{A}$ defined in Sec.~\ref{sec:Entanglement} and averaging over translated strip subsystem origins and quantum trajectories, we obtain
\begin{equation}
    \overline{F}(\ell_y)=\sum_{r_y=1}^{N_y-1} \min(r_y,\ell_y,N_y-r_y)\,\overline{C_G}(r_y).
    \label{eq:FKernel}
\end{equation}
We argue that $\overline{C_G}$ must follow a power-law behavior $\overline{C_G} \propto D(r_y)^{-\beta}$ with $\beta=2$ in the long-time limit to be consistent with the logarithmic scaling of $\overline{F}(\ell_y)$ [Eq.~\eqref{eq:CFTScaling}]. For simplicity, let us consider the limit $N_y \rightarrow \infty$, where the chord lengths simplify to $D(r_y) = r_y$ and $D(\ell_y) = \ell_y$. In this limit, the dominant contribution to the sum in Eq.~\eqref{eq:FKernel} comes from the range $1\leq r_y \leq \ell_y$, where $\min(r_y,\ell_y,N_y-r_y) = r_y$. Therefore, the power-law behavior $\overline{C_G} \propto D(r_y)^{-\beta}$ with $\beta=2$ and the logarithmic scaling of $\overline{F}(\ell_y)$ imply each other. 

In fact, a more careful analysis shows that Eqs.~\eqref{eq:FKernel} and \eqref{eq:CFTScaling} imply the asymptotic behavior of $\overline{C_G}$
\begin{align}
    \overline{C_G}(r_y) = \frac{k}{2\pi^2}D(r_y)^{-2} + {\rm subleading~terms},
    \label{eq:TwoWallPrediction}
\end{align}
at large $D(r_y)$ in the long-time limit of our dynamical system. Recall that $k$ is the coefficient in the logarithmic scaling of $\overline{F}(\ell_y)$ [Eq.~\eqref{eq:CFTScaling}]. To derive the prefactor $\frac{k}{2\pi^2}$, one must carefully include the contributions to the sum in Eq.~\eqref{eq:FKernel} from both the region with small $r_y$ and the region with small $N_y-r_y$. In Sec.~\ref{sec:Entanglement}, we showed that the fitted value of $k$ is very close to 1. In Fig.~\ref{fig:Correlations}(c), we show that the numerically calculated value of $2\pi^2 D(r_y)^2\, \overline{C_G}$ (in the long-time limit of the dynamics) indeed asymptotes toward $k=1$ for large separation $r_y$, confirming the asymptotic behavior Eq.~\eqref{eq:TwoWallPrediction} without any fitting. 

Notice that for the ground state of the $1+1$D free-fermion CFT, the equal-time density correlation also follows the asymptotic behavior $\frac{1}{2\pi^2}D(r_y)^{-2}$ for two density operators separated by $r_y$, matching the long-time behavior Eq.~\eqref{eq:TwoWallPrediction} of the DTDMs. When we split the free-fermion CFT into its chiral and anti-chiral sectors, each sector contributes half, i.e., $\frac{1}{4\pi^2}D(r_y)^{-2}$, of the total density correlation. Similar splitting occurs in our dynamical system. In our numerical simulation of the dynamics, $\overline{C_G}$ also receives equal contributions from the two domain walls. More precisely, we observe that the $\frac{k}{2\pi^2}D(r_y)^{-2}$ scaling of $\overline{C_G}$ can be traced back to the terms in the summation Eq.~\eqref{eq:SquaredCorrelator} where both $x$ and $x'$ are close to the same domain wall, and the two domain walls contribute equally. This observation further supports the striking similarity between the long-time limit of each dynamical topological domain wall and the ground state of the chiral (or anti-chiral) sector of a $1+1$D free-fermion CFT.

\subsection{Slow purification and the Lyapunov spectrum}
\label{sec:SlowDynamics}
\label{sec:Purification}

Having studied the long-time behavior of the dynamical topological domain walls, we next turn to the system's purification dynamics, focusing on how its total entropy decays (i.e., how it purifies) as it evolves from a maximally mixed initial state.
In Ref.~\cite{BhuiyanPanJian2026}, we provided preliminary numerical evidence that the system's total entropy decays algebraically in the presence of the dynamical topological domain walls (as opposed to exponentially when the topological domain wall is absent). Reference~\cite{XiaoKawabata2026} has observed similar behaviors in a postselected analog of our dynamics. However, we caution that the postselection of measurement outcomes in Ref.~\cite{XiaoKawabata2026} renders the dynamics closer to the imaginary-time evolution of a fixed non-disordered Hamiltonian. In contrast, without postselection, our adaptive dynamics naturally experiences randomness in the measurement outcomes. In the following, we elucidate how the dynamical topological domain wall causes the slow algebraic decay of the entropy in our adaptive dynamics (without postselection). As before, we fix $N_x=20$ and evolve for $t=2N_y$ cycles.

\begin{figure*}[!t]
    \StackedFigure
    \centering
    \includegraphics[width=\textwidth]{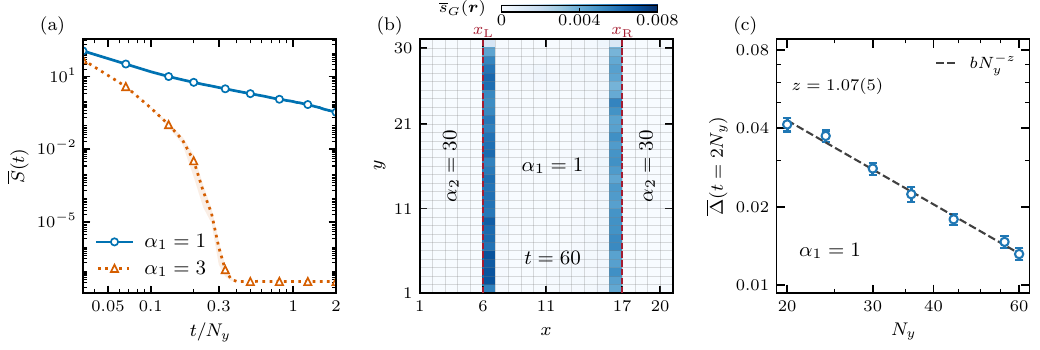}
    \caption{\textbf{(a)} Trajectory-averaged full-system entropy for $N_y=30$ and $\alpha_1=1,3$, starting from the maximally mixed state with measurements throughout the slab and exterior. The plateau for $\alpha_1=3$ is a numerical floor. \textbf{(b)} Trajectory-averaged entanglement contour of the full system~[Eq.~\eqref{eq:Contour}] at $t=60$ for $\alpha_1=1$. Red dashed lines mark the topological domain walls. \textbf{(c)} Trajectory-averaged finite-time Lyapunov gap [Eq.~\eqref{eq:LyapunovFromOccupations}] at $t=2N_y$ for $\alpha_1=1$. The black dashed line is a fit to $bN_y^{-z}$, giving $z=1.07(5)$.} 
    \label{fig:Purification}
\end{figure*}

In Fig.~\ref{fig:Purification}(a), we track the average entropy $\overline{S}(t)$ of the \textit{full system}. This entropy can be viewed as the entanglement entropy between our physical system and a reference system that is maximally entangled with the physical system at $t=0$. When the domain walls are topologically trivial, i.e., $\alpha_1=3$, the entropy $\overline S(t)$ decays exponentially and reaches a numerical floor set by the machine precision after $\mathcal{O}(1)$ cycles. However, in the presence of topological domain walls, i.e., $\alpha_1=1$, the entropy $\overline{S}(t)$ decays algebraically over the simulated window. To locate the region with slow entropy relaxation, we spatially resolve the total entropy of the system using the trajectory-averaged entanglement contour [Eq.~\eqref{eq:Contour}], with the subsystem taken to be the full system. We find that the contour is localized at the two domain walls [Fig.~\ref{fig:Purification}(b)], implying that the $1+1$D DTDMs carry the slowly relaxing entropy and prevent exponentially fast purification of the total entropy. As a comparison, both the $1+1$D DTDMs studied here and their $0+1$D counterparts identified in Ref.~\cite{PanShapourianJian2025} carry long-lived entanglement during purification. This entanglement decays to zero as $t \to \infty$ in the $1+1$D case, whereas it persists indefinitely in the $0+1$D case. Slow purification of boundary modes has also been reported in other topological monitored free-fermion dynamical systems, where it was connected to low-lying Lyapunov spectra~\cite{Oshima2025,XiaoKawabata2026}, which we examine next.

Along each quantum trajectory, the system's state is set by the total circuit evolution operator $\hat V_{\rec}$, the ordered product of Kraus operators along the trajectory (Sec.~\ref{sec:Trajectories}). For the maximally mixed initial state, we have
\begin{align}
    \hat\rho_{\rec}(t)&=\frac{\hat V_{\rec}\hat V_{\rec}^\dagger}{\Tr[\hat V_{\rec}\hat V_{\rec}^\dagger]}\propto e^{-2t\hat{\mathcal H}_{\rec}(t)},\nonumber\\
    &\hat{\mathcal H}_{\rec}(t)=-\frac{1}{2t}\log\big(\hat V_{\rec}\hat V_{\rec}^\dagger\big),
    \label{eq:PurificationKrausProduct}
\end{align}
where $\hat{\mathcal H}_{\rec}(t)$ is the Lyapunov operator. Note that while $\hat V_{\rec}$ need not conserve charge, since it includes gain and loss operations, $\hat V_{\rec}\hat V_{\rec}^\dagger$ does. Combining this fact with the Gaussianity of the evolution allows us to express the many-body Lyapunov operator as a charge-conserving quadratic operator $\hat{\mathcal H}_{\rec}(t)=\hat{\bs c}^\dagger h_{\rec}(t)\,\hat{\bs c}$, up to an additive constant. Here, $\hat{\bs c}^\dag$ and $\hat{\bs c}$ are vectors of fermion creation and annihilation operators labeled by the location and orbital indices $(\bfr,\mu)$. The single-particle Lyapunov operator $h_{\rec}(t)$ is a matrix that can then be accessed through the covariance matrix of the full system via $G(t)=[\mathbf{1}+e^{2th_{\rec}(t)}]^{-1}$. The eigenvalues $\nu_a(t)$ of $G(t)$ therefore determine the finite-time Lyapunov exponents, defined by $\mr{spec}[h_{\rec}(t)]=\{\gamma_a(t)\}_a$, and the associated Lyapunov gap $\Delta(t)$:
\begin{equation}
    \gamma_a(t)=\frac{1}{2t}\log\frac{1-\nu_a(t)}{\nu_a(t)},\qquad
    \Delta(t)=\min_a|\gamma_a(t)|.
    \label{eq:LyapunovFromOccupations}
\end{equation}
Comparing with Eq.~\eqref{eq:EntanglementEnergies}, $2t\gamma_a(t)$ plays the role of the entanglement energies, now for the full system rather than a subsystem. We expect the distribution of the low-lying part of the spectrum $\{\gamma_a(t)\}$ and the gap $\Delta(t)$ to be $t$-independent once $t\sim\mathcal{O}(N_y)$.

Equation~\eqref{eq:PurificationKrausProduct} has the form of a thermal state of the Lyapunov operator at inverse temperature $2t$, so each trajectory is filtered toward the state that fills Lyapunov modes with negative exponents and leaves the ones with positive exponents empty.
Modes with $|\gamma_a|\gg1/t$ are then nearly pure, and the entropy remaining at time $t$ is carried mainly by the modes with small $|\gamma_a|t$. A Lyapunov gap that stays finite as the system grows therefore gives exponential purification at rate $2\Delta$, whereas slow purification requires a gap that closes with system size. Since the measurements in our adaptive circuit are projective, $\hat V_{\rec}\hat V_{\rec}^\dagger$ is rank deficient. Hence, some Lyapunov exponents are infinite, and the corresponding modes are exactly pure. In Secs.~\ref{sec:Entanglement} and~\ref{sec:WallChirality}, we have shown that the long-time state in each quantum trajectory describes a Chern insulator in the center slab and a trivial insulator in the exterior. The domain walls between these regions support states that resemble chiral free-fermion CFT ground states. Hence, it is tempting to compare the finite part of the Lyapunov spectrum with that of a Hamiltonian for a system in which a Chern insulator and a trivial insulator are separated by domain walls. Below, we focus on the lowest-lying part of the Lyapunov spectrum, which controls the slow purification dynamics.

Numerically, we find that the trajectory-averaged Lyapunov gap at fixed aspect ratio $t/N_y=2$ closes as $\overline\Delta\propto N_y^{-z}$ with $z=1.07(5)$ [Fig.~\ref{fig:Purification}(c)], close to the dynamical exponent $z=1$ of a chiral free fermion. Since the residual entropy sits at the walls [Fig.~\ref{fig:Purification}(b)], these smallest-magnitude Lyapunov exponents are carried by the chiral DTDMs, the dynamical analog of the gapless edge states of an equilibrium topological Hamiltonian.   Since these exponents scale as $1/N_y$, the walls purify on a timescale of order $N_y$. We therefore expect the universal entanglement scaling of the DTDMs in the pure-state dynamics to converge within $\mathcal{O}(N_y)$ cycles, consistent with our choice $t=2N_y$ in Secs.~\ref{sec:Entanglement}--\ref{sec:Correlations}.

\subsection{Trajectory-averaged dynamics}
\label{sec:TrajectoryAveraged}

Finally, we ask what remains of the chiral DTDMs when the measurement outcomes are not recorded. Without accessing the measurement outcomes, one should consider the evolution of the system's density matrix $\overline{\hat\rho}$ averaged over all possible quantum trajectories. The quantum channel induced by our adaptive circuit governs this evolution. The measurements defined by the Kraus operators in Eqs.~\eqref{eq:KrausLower} and \eqref{eq:KrausUpper} induce the quantum channels acting on $\overline{\hat\rho}$:
\begin{align}
    \mc E_{i,-}(\overline{\hat\rho})&=\hat{\mc N}_{i,-}\overline{\hat\rho}\,\hat{\mc N}_{i,-}+\hat\chi^\dagger_{i,-}\overline{\hat\rho}\,\hat\chi_{i,-},\nonumber\\
    \mc E_{i,+}(\overline{\hat\rho})&=(\hat\Id-\hat{\mc N}_{i,+})\overline{\hat\rho}\,(\hat\Id-\hat{\mc N}_{i,+})+\hat\chi_{i,+}\overline{\hat\rho}\,\hat\chi^\dagger_{i,+}.
    \label{eq:AveragedChannel}
\end{align}
It turns out that these quantum channels are {\it Gaussian} as they can be rewritten as
\begin{align}
    & \mc E_{i,-}({\bhrho}) = {\rm tr}_{(i,-)}({{\bhrho}}) \, \otimes \kett{1}\braa{1}_{(i,-)}, \nonumber \\ ~~  
    &\mc E_{i,+}({\bhrho}) = {\rm tr}_{(i,+)}({{\bhrho}}) \, \otimes \kett{0}\braa{0}_{(i,+)},
\end{align}
where ${\rm tr}_{(i,\sigma = \pm)}$ denotes the partial trace over the OW mode with index $(i,\sigma)$ and $\kett{s}\braa{s}_{(i,\sigma)}$ denotes the state of the OW mode $(i,\sigma)$ with occupation $s=0,1$. For our dynamical systems, we always have $s = (1-\sigma)/2$. 

Calculating the two-point functions using the average density matrix $\bhrho$ directly yields the covariance matrix $\overline{G}$ averaged over all quantum trajectories. $\overline{G}$ faithfully represents the density matrix $\bhrho$ when $\bhrho$ is Gaussian. The Gaussian quantum channel $\mc E_{(i,\sigma)}$ updates the average covariance matrix $\overline G$ according to
\begin{align}
    \overline G'= \left(1-P_{(i,\sigma)}\right) \overline{G} \left(1-P_{(i,\sigma)}\right) + \frac{1-\sigma}{2} P_{(i,\sigma)}.
    \label{eq:GbarUpdate}
\end{align}
Equivalently, Eq.~\eqref{eq:GbarUpdate} follows from computing the expected value of the trajectory-resolved update rule Eq.~\eqref{eq:CovarianceUpdate} with respect to measurement outcomes.

Therefore, we can numerically simulate the evolution of the trajectory-averaged density matrix $\bhrho$ under our adaptive circuit using this update rule for $\overline{G}$. By combining the effects of all the quantum channels within a cycle of the adaptive circuit, we can show that the evolution of the average covariance matrix takes the form 
\begin{align}
    \overline G_{t+1} = {\cal V} \overline G_{t} {\cal V}^\dag + \Upsilon ,
\end{align}
where $\overline G_t$ denotes the average covariance matrix at circuit cycle $t$ and 
\begin{align}
    {\cal V} = \prod_{(i,\sigma)} \left(1-P_{(i,\sigma)}\right).
\end{align}
The term $\Upsilon$ can be derived from combining the update rules in Eq.~\eqref{eq:GbarUpdate}---we keep its expression implicit, as it does not affect the subsequent discussion~\footnote{Since the choice of ordering does not affect the long-time dynamics, we have left the ordering of elementary operations in the definitions of $\mathcal{V}$ and $\Upsilon$ implicit.}.

\begin{figure}[!t]
    \StackedFigure
    \centering
    \includegraphics[width=\columnwidth]{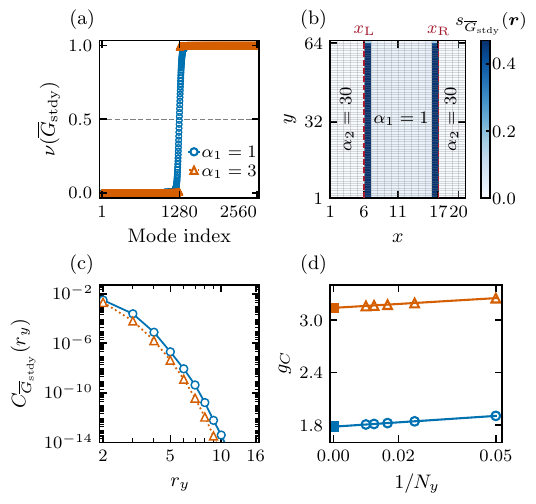}
    \caption{\textbf{(a)} Occupation spectrum of the trajectory-averaged steady state for $\alpha_1=1$ (blue circles) and $3$ (brown triangles). \textbf{(b)} Entropy contour of the trajectory-averaged steady state for $\alpha_1=1$. Dashed lines mark the walls. \textbf{(c)} Squared correlator of the trajectory-averaged steady state for $\alpha_1=1$ and $3$. Panels (a)--(c) use $N_y=64$ and 128 cycles. \textbf{(d)} Relaxation rate $g_C$ [Eq.~\eqref{eq:relaxation_rate}] vs $1/N_y$, with linear extrapolations to $N_y\to\infty$. The legend in (a) applies to (c) and (d) as well.}
    \label{fig:TrajectoryAveraged}
\end{figure}

The average covariance matrix has a steady state $\overline G_{\rm stdy}$ satisfying the equation $ \overline G_{\rm stdy} = {\cal V} \overline G_{\rm stdy} {\cal V}^\dag + \Upsilon$, while the deviation from the steady state $\delta \overline G_t \equiv \overline G_t - \overline{G}_{\rm stdy}$ follows $\delta \overline G_{t+1} = {\cal V} \delta \overline G_{t} {\cal V}^\dag$. Therefore, a generic deviation from $\overline G_{\rm stdy}$ decays as 
\begin{align}
    \|\delta\overline G_t\|\sim e^{-g_C t}. 
    \label{eq:relaxation_rate}
\end{align}
Here, the relaxation rate $g_c$ is given by the negative of the logarithm of the maximal-magnitude eigenvalue of ${\cal V}\otimes {\cal V}^*$.

In the following, we study how $\bhrho$ evolves starting from a maximally mixed initial state on a system of size $N_x \times N_y = 20 \times 64$. We first examine the steady state of $\bhrho$ through the spectrum of the eigenvalues $\nu_a$ of its covariance matrix $\overline G_{\rm stdy}$, shown in Fig.~\ref{fig:TrajectoryAveraged}(a). For both $\alpha_1 = 1$ and $\alpha_1 = 3$, the steady state of $\bhrho$ closely reproduces the local properties of the circuit's pure target state in each region away from the domain walls. Accordingly, most eigenvalues $\nu_a$ cluster near $0$ and $1$. Recall that the eigenvalues of the covariance matrix of a pure Gaussian state can only be 0 or 1. For $\alpha_1 = 1$, however, the target states in the slab bulk and exterior ($\alpha_2 = 30$) have different topologies, and a small subset of eigenvalues spans $(0,1)$, indicating ``mixedness'' in the steady state associated with the topologically nontrivial domain walls in circuit dynamics. The so-called purity gap \cite{BardynEtAl2013}, namely the gap between eigenvalues $\nu_a$ clustered around 0 and 1, is absent. By contrast, for $\alpha_1 = 3$, the target states in different regions share the same topology, and all eigenvalues cluster near $0$ and $1$, separated by a finite purity gap.

Figure~\ref{fig:TrajectoryAveraged}(b) shows the steady-state entropy contour $s_{\overline G_{\rm stdy}}(\bfr)$, obtained by applying Eq.~\eqref{eq:Contour} to $\overline G_{\rm stdy}$. This entropy contour quantifies the spatial distribution of the mixedness in the system. For $\alpha_1 = 1$, most of the steady-state entropy is concentrated near the topological domain wall, where the contour rises above a small but nonzero background. This background arises from truncating the OW modes to a finite range. Even without a domain wall, the circuit's target state is no longer an exact common eigenstate of the number operators associated with the truncated OW modes. At late times, the states in individual trajectories therefore deviate slightly from the local target state and from one another, so their average $\bhrho$ retains some mixedness even far from the wall. The topological domain wall produces a localized enhancement of the entropy contour above the background set by truncation.

Next, we examine the steady-state density correlation $C_{\overline G_{\rm stdy}}$, obtained by applying Eq.~\eqref{eq:SquaredCorrelator} to $\overline G_{\rm stdy}$. For both $\alpha_1 = 1$ and $3$, this correlation decays exponentially [Fig.~\ref{fig:TrajectoryAveraged}(c)]. In contrast, averaging the density correlations computed within each individual trajectory yields $\overline{C_G}$, which decays as a power law in the presence of a topological domain wall ($\alpha_1 = 1$), as discussed in Sec.~\ref{sec:Correlations}. This difference can be understood in terms of phase cancellation. Repeated noncommuting measurements near the domain wall can generate random, trajectory-dependent phases in the two-point functions $G_{(\bfr,\mu),(\bfr',\mu')}$. These phases can lead to cancellations when $G$ is averaged over trajectories. Conceptually similar cancellations occur when two-point functions in disordered equilibrium systems are averaged over disorder realizations.

Moreover, we study the relaxation rate $g_C$ [Eq.~\eqref{eq:relaxation_rate}] of the average density matrix $\bhrho$. Figure~\ref{fig:TrajectoryAveraged}(d) shows the relaxation rates for both $\alpha_1=1$ and $3$ with various $N_y$. Extrapolating $g_C$ to $N_y \rightarrow \infty$ yields $g_C \approx 3.1$ for $\alpha_1 =3$ and $g_C \approx 1.8$ for $\alpha_1 =1$. In both cases, the relaxation rate remains finite when $N_y \rightarrow \infty$, indicating that the average density matrix $\bhrho$ relaxes to the steady state in $\mathcal{O}(1)$ circuit cycles. This again contrasts with the behavior of the dynamical topological domain wall ($\alpha_1 =1$) in individual trajectories where the purification dynamics is algebraically slow (see Sec.~\ref{sec:Purification}) and the Lyapunov gap vanishes as $N_y\rightarrow \infty$. For the dynamics of the average density matrix $\bhrho$, the topological domain wall closes the purity gap in the steady state, but does not lead to a vanishing relaxation rate.

A no-go result for Gaussian Lindbladians~\cite{Mao2025} provides a point of comparison for the Gaussian channel induced by our adaptive circuit. The theorem rules out robust dissipation-free edge modes protected by bulk topology in a Gaussian Lindbladian system that has a finite bulk dissipation gap and a bounded spectrum. Our observations are consistent with this result. The trajectory-averaged density matrix $\bhrho$ relaxes at a finite rate under the Gaussian channel induced by our adaptive circuit. Its steady-state entropy contour reveals pronounced mixedness concentrated near the topological domain walls of the adaptive circuit [Fig.~\ref{fig:TrajectoryAveraged}(b)]. Thus, the average density matrix shows no dissipation-free domain-wall modes.

\subsection{Physical picture for the long-time behavior of the dynamical topological domain wall}
\label{sec:picture}
The adaptive circuit steers the center slab and exterior toward local target states describing a Chern insulator and a trivial insulator, respectively. Both can be viewed as ground states of local Hamiltonians. Near the topological domain wall, no target state is specified, and incompatible measurements generate substantial randomness. Nevertheless, in each quantum trajectory, we observe in Secs.~\ref{sec:Entanglement}--\ref{sec:Correlations} that the wall's long-time behavior approaches that of a (disordered) topological domain wall in the ground state of a Hamiltonian interpolating between the two types of insulators.

We propose a physical picture for this surprising finding by viewing the adaptive circuit as a quantum cooling algorithm. Away from the domain wall, the circuit detects excitations above the local target ground states and removes them through local feedforward operations. Particle and hole excitations near the wall can propagate into the slab bulk or exterior~\footnote{The details of particle and hole propagation depend on the specific implementation of the domain wall in the adaptive protocol.}. The domain wall is thus cooled by proximity, even though the circuit does not explicitly target a specific domain-wall ground state. In this picture, the entire system approaches zero temperature at long times, reaching the ground state of a Hamiltonian with a topological domain wall separating the Chern and trivial insulators.

This cooling picture applies to individual quantum trajectories. Measurement-outcome randomness due to incompatible measurements near the domain wall and the truncation of the OW modes produces different states across trajectories. Averaging over these states is conceptually similar to averaging over ground states of different disorder realizations and can obscure important topological phenomena. Consequently, the novel logarithmic entanglement scaling and power-law squared correlation function on the topological domain wall are observed only in individual quantum trajectories and not in the trajectory-averaged density matrix.

% ------------------------------ SECTION IV ----------------------------
% ------------------------------ SECTION IV ----------------------------
\section{Conclusions and Outlook}
\label{sec:Conclusions}

We studied DTDMs on spacetime domain walls between topologically distinct dynamical phases of $2+1$D free-fermion dynamics, realized by topological adaptive circuits that steer a center slab toward a Chern insulator and its exterior toward a trivial insulator. Although the circuit targets no particular state at the domain walls and subjects them to spatiotemporal randomness, each wall hosts a $1+1$D chiral DTDM. In the long-time limit of each quantum trajectory, the entanglement entropy and charge variance along the walls grow logarithmically with prefactors $c\approx k\approx1$. These prefactors surprisingly match the central charge and the level of the $U(1)$ Kac--Moody algebra of a free-fermion CFT at zero temperature. The density correlations decay algebraically with $\beta\approx2$ and approach the overall amplitude $1/2\pi^2$, which also agrees with the free-fermion CFT. Using the modular commutator, we confirm the chirality of the DTDMs on each domain wall and extract the chiral central charge $c_-=\pm1$. In addition to the novel long-time behavior, the DTDMs also feature slow purification dynamics, with a Lyapunov gap that closes as $N_y^{-z}$ with dynamical exponent $z\approx1$. This slow purification of the DTDMs sharply contrasts with the dynamics in the regions away from the domain wall, where the purification timescale is $\mathcal{O}(1)$.

All these interesting long-time features and purification dynamics are intrinsic to individual quantum trajectories:
They appear only in quantities nonlinear in the density matrix in individual quantum trajectories.
Directly averaging the density matrix over quantum trajectories yields a short-range-correlated mixed state in the long-time limit. The timescale of the relaxation of this averaged state is $\mathcal{O}(1)$. To understand why the adaptive dynamics reproduces the long-wavelength physics of the equilibrium CFT ground state at long times, we have proposed that the circuit acts as a local cooling algorithm, which removes excitations in the bulk and thereby cools the walls by proximity. Topological phases of adaptive dynamics can therefore protect chiral modes on their spacetime boundaries against Born-rule randomness.

Our work leaves several open directions. It would be of interest to extend our construction to $2+1$D topological dynamical phases without charge conservation, where an analogous adaptive circuit would target a chiral $p+\ii p$ superconductor~\cite{BhuiyanPanJian2026}. In this case, we expect that the domain walls would host $1+1$D chiral Majorana DTDMs with half the entanglement found here, while point defects may host $0+1$D modes analogous to vortex-bound Majorana zero modes~\cite{ReadGreen2000,TeoKane2010}.

An analytic description of the chiral DTDMs studied here also remains open. Based on previous studies of free-fermion dynamics with charge conservation~\cite{JianBauerKeselmanLudwig2022,PoboikoEtAl2023,PoboikoGornyiMirlin2024,FavaEtAl2024,StarchlFischerSieberer2025}, the behavior of the $1+1$D chiral DTDMs may admit a field-theoretic description in terms of a nonlinear sigma model in the Altland–Zirnbauer symmetry class AIII.
Deriving this nonlinear sigma model, including any topological term, from the microscopic adaptive circuit, and understanding how the long-time behavior and the slow purification dynamics emerge are natural next questions.

Finally, we comment on experimental prospects. On current qubit platforms, encoding fermions typically requires either nonlocal transformations or additional overhead~\cite{BravyiKitaev2002, VerstraeteCirac2005}. Recent progress on fermionic quantum processors, with fermionic degrees of freedom at the hardware level~\cite{GonzalezCuadra2023,Ott2025,Schuckert2024,zatelli2024robust}, may enable a native realization in the future. On such platforms, occupation measurements could be made strictly finite-range and without additional overhead, and the fresh-ancilla feedforward correction maps naturally onto mid-circuit reset. Because the protocol is strictly Gaussian, it admits efficient classical simulation, which enables systematic benchmarking of gate and measurement fidelities before implementation. Our adaptive circuit is thus a step toward realizing controllable chiral modes on programmable quantum hardware.

\section*{Acknowledgement}

We used OpenAI Codex and Anthropic Claude to assist with manuscript drafting and editing, as well as writing code for numerical analysis, plotting, figure preparation, and exploring analytical arguments. The authors take responsibility for the final content, including the scientific analysis, conclusions, and references.
A.B. was supported by a Cornell fellowship from the Cornell University Graduate School. H.P. is supported by startup funds at the University of Florida. C.-M.J. is supported by the Alfred P. Sloan Foundation through a Sloan Research Fellowship.
The authors acknowledge UFIT Research Computing for providing computation resources and support that have contributed to the research results reported in this publication.

\appendix
\section{Bootstrap uncertainty estimates}
\label{app:Bootstrap}

To estimate uncertainties in fitted parameters, we treat each Born-rule trajectory as an independent sample. For subsystem quantities, we also average over subsystem origins within each trajectory. We then resample trajectories with replacement within each parameter set and system size, preserving the number of trajectories in each group, recompute the ensemble means, and repeat the original fit. For fits across multiple system sizes, each resampling is followed by a joint fit using all included sizes. We repeat this procedure 2,000 times and quote the standard deviation of the fitted parameters as the fit uncertainty. The trajectory-averaged channel of Sec.~\ref{sec:TrajectoryAveraged} is computed exactly and carries no sampling error.

\section{Finite-size effects and logarithmic scaling}
\label{app:SupplementaryNumerics}

\begin{figure*}[htb!]
    \centering
    \includegraphics[width=\textwidth]{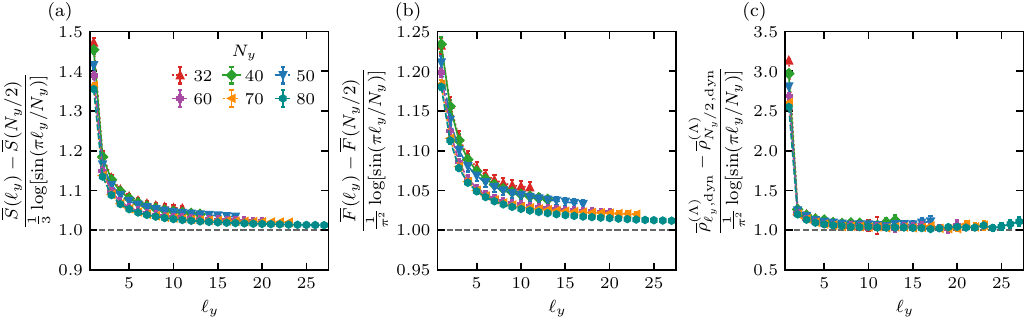}
    \caption{Half-strip-subtracted quantities divided by their free-fermion CFT predictions for $\alpha_1=1$, $N_x=20$, and $t=2N_y$. \textbf{(a)} Entanglement entropy and \textbf{(b)} charge variance, Eq.~\eqref{eq:RatioPrediction}. \textbf{(c)} Window-averaged density of entanglement energies, Eq.~\eqref{eq:WindowDensityRatio}. Dashed lines mark the reference value one. Error bars denote sample-wise SEM per $N_y$.}
    \label{fig:EntropyRatio}
\end{figure*}

To examine the small excess of $c$ and $k$ over one without choosing a fit window, we divide the half-strip-subtracted averages by the predictions of Eq.~\eqref{eq:CFTScaling} with $c=k=1$. If finite-size effects are the cause of the excess, then the ratios
\begin{align}
    &\frac{\overline S(\ell_y)-\overline S(N_y/2)}{\frac13\log\sin(\pi\ell_y/N_y)},\nonumber\\[5pt]
    &\frac{\overline F(\ell_y)-\overline F(N_y/2)}{\frac{1}{\pi^2}\log\sin(\pi\ell_y/N_y)}
    \label{eq:RatioPrediction}
\end{align}
should tend to one with increasing subsystem size. In Fig.~\ref{fig:EntropyRatio}(a,b), we see that both ratios approach one from above, with deviations that decrease with $\ell_y$ and with $N_y$, consistent with the finite-size corrections to $\overline{C_G}$ in Sec.~\ref{sec:Correlations}.

Next, we test the origin of this logarithmic scaling. As discussed in Sec.~\ref{sec:Entanglement}, the logarithmic scaling of the trajectory-averaged entanglement entropy and charge variance follows from a flat spectral density of entanglement energies whose height grows logarithmically with subsystem size [Eq.~\eqref{eq:rho_ansatz}]. To test this growth directly, we define the window-averaged density
\begin{equation}
    \overline{\rho}_{\ell_y,\mathrm{dyn}}^{(\Lambda)}
    \equiv \frac{1}{2\Lambda}\int_{-\Lambda}^{\Lambda}
    \overline{\rho}_{\mathsf{A}(\ell_y),\mathrm{dyn}}(\varepsilon)\,d\varepsilon,
    \label{eq:WindowAveragedDensity}
\end{equation}
where $\Lambda$ is chosen such that the subsystem occupations are restricted to $|2\nu(\varepsilon)-1|<0.99$. Equivalently, $\overline{\rho}_{\ell_y,\mathrm{dyn}}^{(\Lambda)}$ is the mean number of entanglement energies per subsystem in the window, divided by $2\Lambda$. Subtracting the half-strip value and dividing by the free-fermion CFT prediction [Eq.~\eqref{eq:rho_ansatz}, with $\zeta=1$] gives
\begin{equation}
    \frac{\overline{\rho}_{\ell_y,\mathrm{dyn}}^{(\Lambda)}
    -\overline{\rho}_{N_y/2,\mathrm{dyn}}^{(\Lambda)}}
    {\frac{1}{\pi^2}\log\sin(\pi\ell_y/N_y)},
    \label{eq:WindowDensityRatio}
\end{equation}
which tends to one with increasing $\ell_y$ and $N_y$, consistent with finite-size effects [Fig.~\ref{fig:EntropyRatio}(c)]. Like the entanglement entropy and charge-variance ratios, this comparison involves no fitted parameter.

\bibliography{references}

@ARTICLE{BaoChoiAltman2021SymmetryEnriched,
       author = {{Bao}, Yimu and {Choi}, Soonwon and {Altman}, Ehud},
        title = "{Symmetry enriched phases of quantum circuits}",
      journal = {Annals of Physics},
         year = 2021,
        month = dec,
       volume = {435},
          eid = {168618},
        pages = {168618},
          doi = {10.1016/j.aop.2021.168618},
archivePrefix = {arXiv},
       eprint = {2102.09164},
 primaryClass = {cond-mat.stat-mech},
       adsurl = {https://scixplorer.org/abs/2021AnPhy.43568618B}
}

@incollection{potter2022entanglement,
  title = {Entanglement {{Dynamics}} in {{Hybrid Quantum Circuits}}},
  booktitle = {Entanglement in {{Spin Chains}}: {{From Theory}} to {{Quantum Technology Applications}}},
  author = {Potter, Andrew C. and Vasseur, Romain},
  editor = {Bayat, Abolfazl and Bose, Sougato and Johannesson, Henrik},
  year = {2022},
  pages = {211--249},
  publisher = {Springer International Publishing},
  address = {Cham},
  url = {https://doi.org/10.1007/978-3-031-03998-0_9},
  urldate = {2024-04-05},
  isbn = {978-3-031-03998-0},
  langid = {english}
}

@article{SangHsieh2021,
  title = {Measurement-protected quantum phases},
  author = {Sang, Shengqi and Hsieh, Timothy H.},
  journal = {Phys. Rev. Res.},
  volume = {3},
  issue = {2},
  pages = {023200},
  numpages = {9},
  year = {2021},
  month = {Jun},
  publisher = {American Physical Society},
  doi = {10.1103/PhysRevResearch.3.023200},
  url = {https://link.aps.org/doi/10.1103/PhysRevResearch.3.023200}
}

@article{BhuiyanPanJian2026,
  author        = {Bhuiyan, Asadullah and Pan, Haining and Jian, Chao-Ming},
  title         = {Free-Fermion Dynamics with Measurements: Topological Classification and Adaptive Preparation of Topological States},
  journal       = {Physical Review Research},
  publisher     = {American Physical Society},
  volume        = {8},
  number        = {2},
  pages         = {023147},
  year          = {2026},
  doi           = {10.1103/nk38-ygyq},
  eprint        = {2507.13437},
  archivePrefix = {arXiv},
  primaryClass  = {quant-ph}
}

@article{CalabreseCardy2004,
  author  = {Calabrese, Pasquale and Cardy, John},
  title   = {Entanglement Entropy and Quantum Field Theory},
  journal = {Journal of Statistical Mechanics: Theory and Experiment},
  volume  = {2004},
  pages   = {P06002},
  year    = {2004},
  doi     = {10.1088/1742-5468/2004/06/P06002}
}

@book{DiFrancescoMathieuSenechal1997,
  author    = {Di Francesco, Philippe and Mathieu, Pierre and S{\'e}n{\'e}chal, David},
  title     = {Conformal Field Theory},
  publisher = {Springer},
  address   = {New York},
  year      = {1997},
  doi       = {10.1007/978-1-4612-2256-9}
}

@article{Kitaev2006,
  author  = {Kitaev, Alexei},
  title   = {Anyons in an Exactly Solved Model and Beyond},
  journal = {Annals of Physics},
  volume  = {321},
  number  = {1},
  pages   = {2--111},
  year    = {2006},
  doi     = {10.1016/j.aop.2005.10.005},
  eprint  = {cond-mat/0506438},
  archivePrefix = {arXiv},
  primaryClass  = {cond-mat.mes-hall}
}

@ARTICLE{ReadGreen2000,
       author = {{Read}, N. and {Green}, Dmitry},
        title = "{Paired states of fermions in two dimensions with breaking of parity and time-reversal symmetries and the fractional quantum Hall effect}",
      journal = {\prb},
         year = 2000,
        month = apr,
       volume = {61},
       number = {15},
        pages = {10267-10297},
          doi = {10.1103/PhysRevB.61.10267},
archivePrefix = {arXiv},
       eprint = {cond-mat/9906453},
 primaryClass = {cond-mat.mes-hall},
       adsurl = {https://scixplorer.org/abs/2000PhRvB..6110267R}
}

@article{zatelli2024robust,
  title = {Robust Poor Man's {{Majorana}} Zero Modes Using {{Yu-Shiba-Rusinov}} States},
  author = {Zatelli, Francesco and {van Driel}, David and Xu, Di and Wang, Guanzhong and Liu, Chun-Xiao and Bordin, Alberto and Roovers, Bart and Mazur, Grzegorz P. and {van Loo}, Nick and Wolff, Jan C. and Bozkurt, A. Mert and Badawy, Ghada and Gazibegovic, Sasa and Bakkers, Erik P. A. M. and Wimmer, Michael and Kouwenhoven, Leo P. and Dvir, Tom},
  year = {2024},
  month = sep,
  journal = {Nat Commun},
  volume = {15},
  number = {1},
  pages = {7933},
  publisher = {Nature Publishing Group},
  issn = {2041-1723},
  url = {https://www.nature.com/articles/s41467-024-52066-2},
  urldate = {2025-03-21},
  copyright = {2024 The Author(s)},
  langid = {english},
  day = 11
}

@book{NielsenChuang2010,
  author    = {Nielsen, Michael A. and Chuang, Isaac L.},
  title     = {Quantum Computation and Quantum Information},
  edition   = {10th Anniversary},
  publisher = {Cambridge University Press},
  location  = {Cambridge, UK},
  year      = {2010},
  isbn      = {9780521635035},
  doi       = {10.1017/cbo9780511976667}
}

@misc{LiDongLedwithKhalaf2024,
  author        = {Li, Qingchen and Dong, Junkai and Ledwith, Patrick J. and Khalaf, Eslam},
  title         = {Constraints on Real-Space Representations of Chern Bands},
  year          = {2024},
  eprint        = {2407.02561},
  archivePrefix = {arXiv},
  primaryClass  = {cond-mat.str-el},
  doi           = {10.48550/arXiv.2407.02561}
}

@article{SkinnerRuhmanNahum2019,
  title         = {Measurement-{{Induced Phase Transitions}} in the {{Dynamics}} of {{Entanglement}}},
  author        = {Skinner, Brian and Ruhman, Jonathan and Nahum, Adam},
  year          = {2019},
  month         = jul,
  journal       = {Phys. Rev. X},
  volume        = {9},
  number        = {3},
  pages         = {031009},
  doi           = {10.1103/PhysRevX.9.031009}
}

@article{TeoKane2010,
  author  = {Teo, J. C. Y. and Kane, C. L.},
  title   = {Topological Defects and Gapless Modes in Insulators and Superconductors},
  journal = {Physical Review B}, volume = {82}, pages = {115120}, year = {2010},
  doi           = {10.1103/PhysRevB.82.115120}
}

@article{JianBauerKeselmanLudwig2022,
  author  = {Jian, C.-M. and Bauer, B. and Keselman, A. and Ludwig, A. W. W.},
  title   = {Criticality and Entanglement in Nonunitary Quantum Circuits and Tensor Networks of Noninteracting Fermions},
  journal = {Physical Review B}, volume = {106}, pages = {134206}, year = {2022},
  doi     = {10.1103/PhysRevB.106.134206},
  eprint  = {2012.04666}, archivePrefix = {arXiv}
}

@article{FavaEtAl2024,
  author  = {Fava, Michele and Piroli, Lorenzo and Bernard, Denis and Nahum, Adam},
  title   = {Monitored fermions with conserved {$U(1)$} charge},
  journal = {Physical Review Research},
  volume  = {6},
  pages   = {043246},
  year    = {2024},
  doi           = {10.1103/PhysRevResearch.6.043246}
}

@article{PoboikoEtAl2023,
  author  = {Poboiko, I. and P{\"o}pperl, P. and Gornyi, I. V. and Mirlin, A. D.},
  title   = {Theory of Free Fermions under Random Projective Measurements},
  journal = {Physical Review X}, volume = {13}, pages = {041046}, year = {2023},
  doi           = {10.1103/PhysRevX.13.041046}
}

@article{PanShapourianJian2025,
  author  = {Pan, H. and Shapourian, H. and Jian, C.-M.},
  title   = {Topological Modes in Monitored Quantum Dynamics},
  journal = {Physical Review B}, volume = {112}, pages = {144301}, year = {2025},
  eprint  = {2411.04191}, archivePrefix = {arXiv},
  doi           = {10.1103/r8mr-zbx1}
}

@article{XiaoKawabata2026,
  author  = {Xiao, Z. and Kawabata, K.},
  title   = {Symmetry and Topology of Monitored Quantum Dynamics},
  journal = {Physical Review B}, volume = {113}, pages = {134307}, year = {2026},
  doi     = {10.1103/1q83-jcbw}, eprint = {2412.06133}, archivePrefix = {arXiv}
}

@article{Oshima2025,
  author  = {Oshima, H. and Mochizuki, K. and Hamazaki, R. and Fuji, Y.},
  title   = {Topology and Spectrum in Measurement-Induced Phase Transitions},
  journal = {Physical Review Letters}, volume = {134}, pages = {240401}, year = {2025},
  eprint  = {2412.11097}, archivePrefix = {arXiv},
  doi           = {10.1103/PhysRevLett.134.240401}
}

@article{StarchlFischerSieberer2025,
  author        = {Starchl, Elias and Fischer, Mark H. and Sieberer, Lukas M.},
  title         = {Generalized {Z}eno Effect and Entanglement Dynamics Induced by Fermion Counting},
  journal       = {PRX Quantum},
  volume        = {6},
  pages         = {030302},
  year          = {2025},
  eprint        = {2406.07673},
  archivePrefix = {arXiv},
  doi           = {10.1103/jppz-vdgn}
}

@article{QiWuZhang2006,
  author  = {Qi, Xiao-Liang and Wu, Yong-Shi and Zhang, Shou-Cheng},
  title   = {Topological quantization of the spin {H}all effect in two-dimensional paramagnetic semiconductors},
  journal = {Phys. Rev. B},
  volume  = {74},
  pages   = {085308},
  year    = {2006},
  doi           = {10.1103/PhysRevB.74.085308}
}

@article{bravyi2004lagrangian,
  author        = {Bravyi, Sergey},
  title         = {Lagrangian Representation for Fermionic Linear Optics},
  journal       = {Quantum Information and Computation},
  volume        = {5},
  number        = {3},
  pages         = {216--238},
  year          = {2005},
  doi           = {10.26421/QIC5.3-3},
  eprint        = {quant-ph/0404180},
  archivePrefix = {arXiv}
}

@article{BiancoResta2011,
  author  = {Bianco, Raffaello and Resta, Raffaele},
  title   = {Mapping topological order in coordinate space},
  journal = {Physical Review B},
  volume  = {84},
  number  = {24},
  pages   = {241106},
  year    = {2011},
  doi     = {10.1103/PhysRevB.84.241106},
  eprint  = {1111.5697},
  archivePrefix = {arXiv},
  primaryClass = {cond-mat.str-el}
}

@article{BardynEtAl2013,
  author  = {Bardyn, C.-E. and Baranov, M. A. and Kraus, C. V. and Rico, E. and {\.I}mamo{\u g}lu, A. and Zoller, P. and Diehl, S.},
  title   = {Topology by Dissipation},
  journal = {New Journal of Physics}, volume = {15}, pages = {085001}, year = {2013},
  doi           = {10.1088/1367-2630/15/8/085001}
}

@article{EislerPeschel2013,
  author        = {Eisler, Viktor and Peschel, Ingo},
  title         = {Free-fermion entanglement and spheroidal functions},
  journal       = {Journal of Statistical Mechanics: Theory and Experiment},
  volume        = {2013},
  number        = {04},
  pages         = {P04028},
  year          = {2013},
  doi           = {10.1088/1742-5468/2013/04/P04028},
  eprint        = {1302.2239},
  archivePrefix = {arXiv},
  primaryClass  = {cond-mat.stat-mech}
}

@article{ChenVidal2014,
  author  = {Chen, Yangang and Vidal, Guifr{\'e}},
  title   = {Entanglement contour},
  journal = {Journal of Statistical Mechanics: Theory and Experiment},
  volume  = {2014},
  number  = {10},
  pages   = {P10011},
  year    = {2014},
  doi     = {10.1088/1742-5468/2014/10/P10011}
}

@article{puente2024quantum,
  author        = {Alcalde Puente, Daniel and Motzoi, Felix and Calarco, Tommaso and Morigi, Giovanna and Rizzi, Matteo},
  title         = {Quantum state preparation via engineered ancilla resetting},
  journal       = {Quantum},
  volume        = {8},
  pages         = {1299},
  year          = {2024},
  doi           = {10.22331/q-2024-03-27-1299},
  eprint        = {2305.08641},
  archivePrefix = {arXiv}
}

@article{roy2020measurement,
  author        = {Roy, Sthitadhi and Chalker, J. T. and Gornyi, I. V. and Gefen, Yuval},
  title         = {Measurement-induced steering of quantum systems},
  journal       = {Phys. Rev. Research},
  volume        = {2},
  pages         = {033347},
  year          = {2020},
  doi           = {10.1103/PhysRevResearch.2.033347},
  eprint        = {1912.04292},
  archivePrefix = {arXiv}
}

@article{song2012bipartite,
  title     = {Bipartite fluctuations as a probe of many-body entanglement},
  author    = {Song, H. Francis and Rachel, Stephan and Flindt, Christian and Klich, Israel and Laflorencie, Nicolas and Le Hur, Karyn},
  journal   = {Phys. Rev. B},
  volume    = {85},
  number    = {3},
  pages     = {035409},
  year      = {2012},
  doi       = {10.1103/PhysRevB.85.035409},
  eprint    = {1109.1001},
  archivePrefix = {arXiv},
  primaryClass  = {cond-mat.str-el}
}

@article{PoboikoGornyiMirlin2024,
  author  = {Poboiko, Igor and Gornyi, Igor V. and Mirlin, Alexander D.},
  title   = {Measurement-Induced Phase Transition for Free Fermions above One Dimension},
  journal = {Physical Review Letters},
  volume  = {132},
  pages   = {110403},
  year    = {2024},
  doi     = {10.1103/PhysRevLett.132.110403}
}

@article{BravyiKitaev2002,
  author  = {Bravyi, Sergey B. and Kitaev, Alexei Yu.},
  title   = {Fermionic Quantum Computation},
  journal = {Annals of Physics},
  volume  = {298},
  pages   = {210},
  year    = {2002},
  doi           = {10.1006/aphy.2002.6254}
}

@article{GonzalezCuadra2023,
  author  = {Gonz{\'a}lez-Cuadra, Daniel and Bluvstein, Dolev and Kalinowski, Marcin and Kaubruegger, Raphael and Maskara, Nishad and Naldesi, Piero and Zache, Torsten V. and Kaufman, Adam M. and Lukin, Mikhail D. and Pichler, Hannes and Vermersch, Beno{\^\i}t and Ye, Jun and Zoller, Peter},
  title   = {Fermionic Quantum Processing with Programmable Neutral Atom Arrays},
  journal = {Proceedings of the National Academy of Sciences},
  volume  = {120},
  pages   = {e2304294120},
  year    = {2023},
  doi           = {10.1073/pnas.2304294120}
}

@article{Ott2025,
  author  = {Ott, Robert and Gonz{\'a}lez-Cuadra, Daniel and Zache, Torsten V. and Zoller, Peter and Kaufman, Adam M. and Pichler, Hannes},
  title   = {Error-Corrected Fermionic Quantum Processors with Neutral Atoms},
  journal = {Physical Review Letters},
  volume  = {135},
  pages   = {090601},
  year    = {2025},
  doi           = {10.1103/zkpl-hh28}
}

@misc{Schuckert2024,
  author        = {Schuckert, Alexander and Crane, Eleanor and Gorshkov, Alexey V. and Hafezi, Mohammad and Gullans, Michael J.},
  title         = {Fermion-Qubit Fault-Tolerant Quantum Computing},
  year          = {2024},
  eprint        = {2411.08955},
  archivePrefix = {arXiv},
  primaryClass  = {quant-ph},
  doi           = {10.48550/arXiv.2411.08955}
}

@article{CerveraLierta2018,
  author  = {Cervera-Lierta, Alba},
  title   = {Exact {I}sing model simulation on a quantum computer},
  journal = {Quantum},
  volume  = {2},
  pages   = {114},
  year    = {2018},
  doi     = {10.22331/q-2018-12-21-114}
}

@article{Mao2025,
  title         = {Absence of Dissipation-Free Topological Edge States in Quadratic Open Fermions},
  author        = {Mao, Liang},
  year          = {2026},
  month         = may,
  journal       = {Quantum Sci. Technol.},
  volume        = {11},
  number        = {2},
  pages         = {025050},
  publisher     = {IOP Publishing},
  doi           = {10.1088/2058-9565/ae6a1a}
}

@article{VerstraeteCiracLatorre2009,
  author  = {Verstraete, Frank and Cirac, J. Ignacio and Latorre, Jos{\'e} I.},
  title   = {Quantum circuits for strongly correlated quantum systems},
  journal = {Phys. Rev. A},
  volume  = {79},
  pages   = {032316},
  year    = {2009},
  doi     = {10.1103/PhysRevA.79.032316}
}

@article{KimShiKatoAlbert2022,
  author  = {Kim, Isaac H. and Shi, Bowen and Kato, Kohtaro and Albert, Victor V.},
  title   = {Chiral Central Charge from a Single Bulk Wave Function},
  journal = {Phys. Rev. Lett.},
  volume  = {128},
  pages   = {176402},
  year    = {2022},
  doi     = {10.1103/PhysRevLett.128.176402}
}

@article{KimShiKatoAlbert2022b,
  author  = {Kim, Isaac H. and Shi, Bowen and Kato, Kohtaro and Albert, Victor V.},
  title   = {Modular commutator in gapped quantum many-body systems},
  journal = {Phys. Rev. B},
  volume  = {106},
  pages   = {075147},
  year    = {2022},
  doi     = {10.1103/PhysRevB.106.075147}
}

@article{ZouShiSorceLimKim2022,
  author  = {Zou, Yijian and Shi, Bowen and Sorce, Jonathan and Lim, Ian T. and Kim, Isaac H.},
  title   = {Modular Commutators in Conformal Field Theory},
  journal = {Phys. Rev. Lett.},
  volume  = {129},
  pages   = {260402},
  year    = {2022},
  doi     = {10.1103/PhysRevLett.129.260402}
}

@article{FanZhangGu2023,
  author  = {Fan, Ruihua and Zhang, Pengfei and Gu, Yingfei},
  title   = {Generalized real-space {C}hern number formula and entanglement hamiltonian},
  journal = {SciPost Phys.},
  volume  = {15},
  pages   = {249},
  year    = {2023},
  doi     = {10.21468/SciPostPhys.15.6.249}
}

@article{HolzheyLarsenWilczek1994,
  author  = {Holzhey, Christoph and Larsen, Finn and Wilczek, Frank},
  title   = {Geometric and renormalized entropy in conformal field theory},
  journal = {Nucl. Phys. B},
  volume  = {424},
  pages   = {443--467},
  year    = {1994},
  doi     = {10.1016/0550-3213(94)90402-2}
}

@article{li2018quantum,
  title         = {Quantum {{Zeno}} Effect and the Many-Body Entanglement Transition},
  author        = {Li, Yaodong and Chen, Xiao and Fisher, Matthew P. A.},
  year          = {2018},
  month         = nov,
  journal       = {Phys. Rev. B},
  volume        = {98},
  number        = {20},
  pages         = {205136},
  publisher     = {American Physical Society},
  doi           = {10.1103/PhysRevB.98.205136}
}

@article{gullans2020dynamical,
  title         = {Dynamical {{Purification Phase Transition Induced}} by {{Quantum Measurements}}},
  author        = {Gullans, Michael J. and Huse, David A.},
  year          = {2020},
  month         = oct,
  journal       = {Phys. Rev. X},
  volume        = {10},
  number        = {4},
  pages         = {041020},
  doi           = {10.1103/PhysRevX.10.041020}
}

@article{agrawal2022entanglement,
  title         = {Entanglement and Charge-Sharpening Transitions in {{U}}(1) Symmetric Monitored Quantum Circuits},
  author        = {Agrawal, Utkarsh and Zabalo, Aidan and Chen, Kun and Wilson, Justin H. and Potter, Andrew C. and Pixley, J. H. and Gopalakrishnan, Sarang and Vasseur, Romain},
  year          = {2022},
  month         = oct,
  journal       = {Phys. Rev. X},
  volume        = {12},
  number        = {4},
  pages         = {041002},
  doi           = {10.1103/PhysRevX.12.041002}
}

@article{barratt2022field,
  title         = {Field Theory of Charge Sharpening in Symmetric Monitored Quantum Circuits},
  author        = {Barratt, Fergus and Agrawal, Utkarsh and Gopalakrishnan, Sarang and Huse, David A. and Vasseur, Romain and Potter, Andrew C.},
  year          = {2022},
  month         = sep,
  journal       = {Phys. Rev. Lett.},
  volume        = {129},
  number        = {12},
  pages         = {120604},
  doi           = {10.1103/PhysRevLett.129.120604}
}

@article{majidy2023critical,
  title         = {Critical Phase and Spin Sharpening in {{SU}}(2)-Symmetric Monitored Quantum Circuits},
  author        = {Majidy, Shayan and Agrawal, Utkarsh and Gopalakrishnan, Sarang and Potter, Andrew C. and Vasseur, Romain and Halpern, Nicole Yunger},
  year          = {2023},
  month         = aug,
  journal       = {Phys. Rev. B},
  volume        = {108},
  number        = {5},
  pages         = {054307},
  publisher     = {American Physical Society},
  doi           = {10.1103/PhysRevB.108.054307}
}

@article{lang2020entanglement,
  title         = {Entanglement Transition in the Projective Transverse Field {{Ising}} Model},
  author        = {Lang, Nicolai and B{\"u}chler, Hans Peter},
  year          = {2020},
  month         = sep,
  journal       = {Phys. Rev. B},
  volume        = {102},
  number        = {9},
  pages         = {094204},
  doi           = {10.1103/PhysRevB.102.094204}
}

@misc{aziz2026measurementonly,
  title         = {Measurement-{{Only Dynamical Phase Transitions}} in {{Spin-1 Chains}}},
  author        = {Aziz, Kemal and Pan, Haining and Pixley, J. H.},
  year          = {2026},
  month         = sep,
  url           = {https://arxiv.org/abs/2609.26941},
  eprint        = {2609.26941},
  archiveprefix = {arXiv},
  doi           = {10.48550/arXiv.2609.26941}
}

@article{jian2023measurementinduced,
  title = {Measurement-Induced Entanglement Transitions in Quantum Circuits of Non-Interacting Fermions: {{Born-rule}} versus Forced Measurements},
  shorttitle = {Measurement-Induced Entanglement Transitions in Quantum Circuits of Non-Interacting Fermions},
  author = {Jian, Chao-Ming and Shapourian, Hassan and Bauer, Bela and Ludwig, Andreas W. W.},
  year = {2023},
  month = feb,
  journal = {arXiv:2302.09094},
  primaryclass = {cond-mat, physics:quant-ph},
  url = {http://arxiv.org/abs/2302.09094},
  urldate = {2023-02-23},
  archiveprefix = {arXiv},
  day = 17
}

@article{LavasaniAlaviradBarkeshli2021,
  title = {Measurement-Induced Topological Entanglement Transitions in Symmetric Random Quantum Circuits},
  author = {Lavasani, Ali and Alavirad, Yahya and Barkeshli, Maissam},
  year = {2021},
  month = mar,
  journal = {Nat. Phys.},
  volume = {17},
  number = {3},
  pages = {342--347},
  publisher = {Nature Publishing Group},
  issn = {1745-2481},
  url = {https://www.nature.com/articles/s41567-020-01112-z},
  urldate = {2024-11-02},
  copyright = {2021 The Author(s), under exclusive licence to Springer Nature Limited},
  langid = {english}
}

@article{lavasani2021topological,
  author = {Lavasani, Ali and Alavirad, Yahya and Barkeshli, Maissam},
  year = {2021},
  month = dec,
  journal = {Phys. Rev. Lett.},
  volume = {127},
  number = {23},
  pages = {235701},
  publisher = {American Physical Society},
  url = {https://link.aps.org/doi/10.1103/PhysRevLett.127.235701},
  urldate = {2024-11-02},
  title = {Topological {{Order}} and {{Criticality}} in {$(2+1)\mathrm{D}$} {{Monitored Random Quantum Circuits}}},
  day = 1
}

@article{fava2023nonlinear,
  title = {Nonlinear {{Sigma Models}} for {{Monitored Dynamics}} of {{Free Fermions}}},
  author = {Fava, Michele and Piroli, Lorenzo and Swann, Tobias and Bernard, Denis and Nahum, Adam},
  year = {2023},
  month = dec,
  journal = {Phys. Rev. X},
  volume = {13},
  number = {4},
  pages = {041045},
  publisher = {American Physical Society},
  url = {https://link.aps.org/doi/10.1103/PhysRevX.13.041045},
  urldate = {2024-09-13},
  day = 8
}

@article{fisher2023random,
  title         = {Random {{Quantum Circuits}}},
  author        = {Fisher, Matthew P. A. and Khemani, Vedika and Nahum, Adam and Vijay, Sagar},
  year          = {2023},
  month         = mar,
  journal       = {Annu. Rev. Condens. Matter Phys.},
  volume        = {14},
  pages         = {335--379},
  doi           = {10.1146/annurev-conmatphys-031720-030658}
}

@article{VerstraeteCirac2005,
  author  = {Verstraete, Frank and Cirac, J. Ignacio},
  title   = {Mapping local {H}amiltonians of fermions to local {H}amiltonians of spins},
  journal = {J. Stat. Mech.},
  year    = {2005},
  pages   = {P09012},
  doi     = {10.1088/1742-5468/2005/09/P09012},
  eprint  = {cond-mat/0508353},
  archivePrefix = {arXiv}
}
\end{document}